\documentclass[10pt,twoside]{article} 

\usepackage{geometry}
\usepackage{amsfonts}
\usepackage{stmaryrd}
\usepackage{amssymb}
\usepackage{euscript}
\usepackage{pstricks}
\usepackage{amsthm}
\usepackage{latexsym,amsmath,amscd}
\usepackage{mathrsfs}
\usepackage{graphicx}
\usepackage{color}
\usepackage{dsfont}
\usepackage{enumerate}
\usepackage{booktabs}

\newcommand{\Keywords}[1]{\par\noindent{\small{\bf Keywords\/}: #1}}
\newcommand{\Class}[1]{\par\noindent{\small{\bf Mathematics Subjects Classification (2010)\/}: #1}}

\newtheorem{theorem}{Theorem}[section]

\newtheorem{lem}{Lemma}[section]
\newtheorem{pro}{Proposition}[section]
\newtheorem{cor}{Corollary}[section]
\newtheorem{rem}{Remark}[section]
\newtheorem{rems}{Remarks}[section]
\newtheorem{ex}{Example}[section]
\newtheorem{defi}{Definition}[section]
\newtheorem{hyp}{Assumption}[section]

\newcommand{\bt}{\begin{theorem}}
\newcommand{\et}{\end{theorem}}
\newcommand{\bl}{\begin{lem}}
\newcommand{\el}{\end{lem}}
\newcommand{\bp}{\begin{pro}}
\newcommand{\ep}{\end{pro}}
\newcommand{\bcor}{\begin{cor}}
\newcommand{\ecor}{\end{cor}}
\newcommand{\lab }{\label }
\newcommand{\bd}{\begin{defi} \rm }
\newcommand{\ed}{\end{defi}}
\newcommand{\brem }{\begin{rem} \rm }
\newcommand{\erem }{\end{rem}}
\newcommand{\brems }{\begin{rems} \rm }
\newcommand{\erems }{\end{rems}}
\newcommand{\bhyp }{\begin{hyp} \rm }
\newcommand{\ehyp }{\end{hyp}}
\newcommand{\bex}{\begin{ex} \rm }
\newcommand{\eex}{\end{ex}}

\newcommand{\ssc}{\subsection}
\newcommand{\sssc}{\subsubsection}

\newcommand{\be}{\begin{equation}}
\newcommand{\ee}{\end{equation}}
\newcommand{\bde}{\begin{displaymath}}
\newcommand{\ede}{\end{displaymath}}
\newcommand{\beq}{\begin{eqnarray*}}
\newcommand{\eeq}{\end{eqnarray*}}
\newcommand{\beqa}{\begin{eqnarray}}
\newcommand{\eeqa}{\end{eqnarray}}
\newcommand{\bel }{\left\{\begin{array}{ll}}
\newcommand{\eel}{\cr \end{array} \right.}

\def\mid{\,|\,}
\def\Mid{\,\big|\,}

\def\AMC{{\cal A}}

\def\bhP{\widehat P}

\def\CVA{{\rm CVA}}

\def\proof{\noindent {\it Proof. $\, $}}
\def\endproof {\hfill $\Box$ \vskip 5 pt }

\def\hat{\widehat }
\def\tilde{\widetilde }

\def\EE{\mathbb{E}_{\mathbb{Q}}}

\def\FG{{\mathbb{G}}}

\def\Q{\mathbb{Q}}

\def\Om{\Omega}

\def\G{{\cal{G}}}

\def\rr{\mathbb{R}}

\def\MM{\mathcal{M}}

\def\I{\mathds{1}}

\usepackage{eso-pic}

\title{{{\Large \bf CVA FOR BILATERAL COUNTERPARTY RISK UNDER ALTERNATIVE SETTLEMENT CONVENTIONS}}\vskip 40 pt}

\author{Cyril Durand \\
Mathematics Department \\
Imperial College\\
London SW7 2AZ, United Kingdom\\ \\
Marek Rutkowski\thanks{The research of M. Rutkowski was supported under Australian
Research Council's Discovery Projects funding scheme (DP120100895).
The authors thank J.P. Laurent for enlightening discussions and comments. 
Forthcoming in IJTAF under the title {\it CVA under Alternative Settlement Conventions and with Systemic Risk.}}
\\ School of Mathematics and Statistics
\\ University of Sydney
\\ NSW 2006, Australia}

\date{\vskip 20 pt March 20, 2012\vskip 20 pt}

\begin{document}
\maketitle

\begin{abstract}
We depart from the usual methods for pricing contracts with the counterparty credit
risk found in most of the existing literature. In effect, typically, these models do not account
for either systemic effects or at-first-default contagion and postulate that the
contract value at default equals either the risk-free value or the pre-default value. We propose instead a fairly general
framework, which allows us to perform effective Credit Value Adjustment (CVA) computations for a
contract with bilateral counterparty risk in the presence of systemic and wrong or right way risks. Our general methodology focuses on
the role of alternative settlement clauses, but it is also aimed to cover various features of margin agreements.
A comparative analysis of numerical results reported in the final section supports our initial conjecture that alternative specifications
of settlement values have a non-negligible impact on the CVA computation for contracts with bilateral counterparty risk.
This emphasizes the practical importance of more sophisticated models that are capable of fully reflecting
the actual features of financial contracts, as well as the influence of the market environment.
\vskip 20 pt
\vskip 20 pt
\Class{91G40,$\,$60J28}
\end{abstract}

\newpage

\section{Introduction} \label{sect1}

The risk that a counterparty cannot
meet its contractual obligations has become the hot subject
of the moment in the realm of finance. Intertwined studies were recently conducted in different directions, especially
the {\it systemic risk} (the risk of the domino effect following the bankruptcy of a major financial
institution, as studied, for instance, by Cont and Moussa \cite{CM2},
the systemic impact of centralized clearing examined by Duffie and Zhu \cite{DZ}, the effects of
 asymmetric information in regard credit risk of securitized products and exposure of banks
 to these products in Gorton \cite{G1}), the {\it liquidity effects} (the risk of fire
 sales, impact of collateral policies of central banks in Gauthier et al. \cite{GLS} or Liberti and Mian \cite{LM}),
  and, last but not least, the {\it counterparty risk} (computations of the {\it Credit Value Adjustment}
 (CVA) and modeling of wrong way risk in Bielecki et al. \cite{BCJZ}, Brigo and Capponi \cite{BC}, Brigo et al. \cite{BCP},
 Cespedes et al. \cite{C},  Cr\'epey et al. \cite{CJZ}, Durand \cite{D}, Gregory \cite{G},
  Lipton and Sepp \cite{LS}, Pykhtin \cite{P}, Elouerkahoui \cite{E}, and Yi \cite{Y1,Y2}, to mention a few).

 While the effects of margin agreements and the intricacies
 of collateral markets (rehypothecation, especially) emerge as a crucial issue from all
  these different perspectives, the regulators are enforcing new constraints on
 liquidity, capital allocation and short term borrowing. They are
 also contemplating new ways of dealing with defaults (most notably, novation) and
 pushing for the development of additional compensation chambers
 for derivatives. Although some support for the introduction of central
  clearing houses for some OTC derivatives  (for instance, credit default swaps) was voiced,
 it is unclear whether the introduction of such
clearing houses would improve counterparty risk management
considered as a whole (see Duffie and Zhu \cite{DZ}). In any case, the study of counterparty risk will
still be necessary for OTC derivatives.
 It is also worthwhile to note that the bankruptcy of investment
 banks, such as Lehman Brothers, shed new light on legal issues related to the termination
 and settlement of derivative contracts in a context of multi-jurisdictions
 and diverse contractual terms, which calls for enhanced harmonization of legal proceedings.
 These issues are beyond the scope of this research, however.

In the aftermath of the recent financial crisis, it is now widely acknowledged that it is crucial to account for both counterparty risk
and systemic risk in the pricing and hedging of financial instruments. In this paper, we depart from
the usual methods for pricing contracts with counterparty risks developed in most of the existing literature. In effect, typically,
these models, first, postulate that the contract's value at default is equal either to the risk-free value
or the pre-default value. Second, they frequently do not take margin agreements into account. Finally, contagion effects are typically neglected. By contrast, we analyze here in some detail the CVA computations under several alternative settlement conventions.
We also quantify the effects of various kinds of margin agreements,
as typically specified in the {\it Credit Support Annex} (CSA) of a contract. Finally,
we propose and analyze a particular Markovian contagion model in which the effects of correlation parameters,
such as the wrong or right way risks or the at-first-default contagion effects, as well as the systemic risk, are taken into account
in an explicit and intuitively appealing manner.

We start by developing in Section  \ref{sect2} a fairly general framework for the CVA computation
of an collateralized or a collateralized contract under the bilateral counterparty risk.
Note that we deal throughout with the bilateral counterparty risk, so that the sign of CVA can be either positive or negative,
in principle. Consequently, there is no need to introduce the terminological distinction between the concepts of CVA and
DVA ({\it Debt Value Adjustment}), as is quite common in those practical applications where the computations
of CVA and DVA are conducted separately, assuming each time the unilateral counterparty risk.
In Section  \ref{sect3},  we focus our attention on the CVA computation for credit default swaps with bilateral counterparty
risk.  We propose there a Markovian model of credit contagion
and we derive algorithms for the CVA computation of a credit default swap in set-ups without and with
systemic risk. Numerical results reported in the final section show significant variations in CDS spreads
for alternative specifications of settlement clauses, levels of systemic risk, and margin agreements.
Selected stylized examples presented in this work support the view that
the CVA computed using the methodology developed in this work may significantly differ from the one obtained through less sophisticated pricing methods. Similar conclusions were obtained in an independent recent research by Brigo et al. \cite{BBM12} who worked in a different framework, however.

\section{Contracts with Bilateral Counterparty Risk} \label{sect2}

The hypothesis of a unilateral counterparty risk has been seen in the past as a practical, albeit somewhat rough, estimate for modeling contracts between a major OECD financial institution and a client (see, for instance, Mashal and Naldi \cite{MN})). However, the realization that even the most prestigious investment banks, such as Lehman Brothers or Merrill Lynch,
could go bankrupt has shattered the foundations for resorting to unilateral models.
The clients of banks are nowadays prone to question such an assumption and are willing to ask for suitable adjustments
of contractual terms in order to gain a better security on their financial instruments, as well as on their collaterals,
in the event of default of a counterparty. This new situation explains why, for instance, the subject of re-hypothecation of collateral has given rise to an active debate between investment banks and hedge funds.

\ssc{Settlement Covenants without Margin Agreements}  \label{sect2.1}

We first recall the basic concepts related to the counterparty risk in a general setting
and introduce the notation for the relevant quantities; we follow here Durand \cite{D}.
By the {\it risk-free contract,} we mean an over-the-counter (OTC) contract between two non-defaultable parties
(one can also refer here to an exchange-traded contract with otherwise identical features).
The {\it counterparty risk} is the risk that the party of an OTC contract may default and fail to meet his contractual obligations.
A contract with {\it unilateral counterparty risk} is an OTC contract between a non-defaultable investor
and a defaultable counterparty.
More generally, by a  contract with {\it bilateral counterparty risk} we mean an OTC contract between
 a defaultable investor and a defaultable counterparty.
We denote by $1$ and $2$ the two parties, which enter into a
contract; they are hereafter referred to  as the `investor' and the
`counterparty'. We write $c^1_t$ and $c^2_t$ to denote the
level of creditworthiness (or credit quality) of each party at time $t$.
Formally, by the {\it creditworthiness } $c^p_t,\, p=1,2$ of the $p$th party at time $t$, we
mean some quantitative measure of the ability at time $t$ of this party to fulfill his obligations, given his financial and business situation.
In practice, as a proxy for the creditworthiness at time $t$ one may take, for instance, the current credit
rating, which can either be based on an internal rating or provided by a specialized rating agency.

In the event of the counterparty's default, the contract is
terminated and immediately settled according to some predetermined rule, which was accepted by the two parties.
For a contract that is liquidly traded after the counterparty's
default, the settlement value could be based on the current {\it market value} of
the contract. Obviously, this would be a rather unusual situation for an OTC contract to be
actively traded after the counterparty default. Therefore, there is a need to specify the settlement value in some
conventional way. For this purpose, we will first examine the concept of the mark-to-market
value of a contract. The determination of the contractual mark-to-market value at
the time of default hinges on specific clauses in the
contract's agreement signed by both parties. It is thus natural to assume that they agreed
upon some convention, denoted hereafter by $i)$, in the event of default.
By the right-continuous with left-hand limits (i.e., c\`adl\`ag) process  $M^i_t$
we denote the conventional mark-to-market value of a contract, as seen by the investor.
The following alternative specifications of the mark-to-market value will be studied in what follows.

\bd \lab{def3.6}
The {\it mark-to-market value} $M^i_t$ ({\it MtM value}, for short) at time $t$ under convention
 $i$) for $i = a, a', b, c, c'$ is the value of a contract with identical promised future cash flows, but which is initiated at time $t$ between: \hfill \break
$a)$  two risk-free companies; \hfill \break
$a')$  the non-defaulting party and a risk-free company; \hfill \break
$b)$   the non-defaulting party and a party with the same creditworthiness at time $t$
 as the initial creditworthiness of the defaulting party; \hfill \break
$c)$  the non-defaulting party and a party with the same creditworthiness at time $t$
as the creditworthiness of the counterparty just prior to default; \hfill \break
$c')$   two companies with the same creditworthiness as the investor and
counterparty enjoyed just prior to default.
\ed

The actual level of the MtM value at the moment of default may also
depend on whether the counterparty or the investor defaulted.
We thus use the notation $M^{i,1}$ and $M^{i,2}$ to emphasize the feature that the MtM value is dependent on
the defaulted name.  The contract's settlement value is typically defined as a particular function of its MtM value
in the event of default. If one of the
counterparties defaults, it is usual to settle the contract as
follows: if the MtM value $M^{i,2}_t$  of the contract at
time $t$ of the counterparty's default is negative for the investor
(which means that the MtM value $-M^{i,2}_t$ is positive for the counterparty) then the counterparty receives the full
value of the contract. Otherwise, that is, when the MtM value $M^{i,2}_t$ is positive, the investor claims the full value
$M^{i,2}_t$, but, typically, she will receive only a portion of this value.
An analogous rule applies in the event of investor's default.

The prevailing market convention, known as the {\it full two-way payment} clause
(or the {\it no-fault rule}) for settlement after early termination of a contract (including  the default event), postulates that
both counterparties are entitled to net payments.
An alternative {\it limited two-way payment} clause permits the non-defaulting party to avoid any net liabilities to
the defaulting party, while claiming any net balance due from the defaulting party.
It was widely criticized for making possible to a non-defaulting counterparty to walk away with a windfall.
The 1992 ISDA Master Agreement (amended in 2002, see \cite{ISDA1}), which is the most commonly used
master contract for OTC derivative transactions, allows both limited and full two-way payments
to be applied, but it states that full two-way payment clause will apply, unless counterparties elect
limited two-way payments. In what follows, we always consider settlement values from the investor's perspective
assuming the full two-way payment rule.
Let us now formally define the {\it settlement value} (also known as the {\it close-out amount})
associated with a particular MtM convention.
We denote $a^+ = \max \, (a,0)$ and $a^- = \max \, (-a, 0)$.

\bd \lab{sv1}
The contract's {\it settlement value} $S^{i,2}_t$  in the case of the
counterparty's default at time $t$ is given by the stochastic process $S^{i,2}_t :=
R^2_t (M^{i,2}_t)^+ - (M^{i,2}_t)^-$ where $R^2_t \in [0,1]$ is the  counterparty's recovery rate.
The contract's {\it settlement value} $S^{i,1}_t$
in the case of the investor's default at time $t$ is given by the stochastic process
$S^{i,1}_t := (M^{i,1}_t)^+ - R^1_t (M^{i,1}_t)^- $ where $R^1_t \in [0,1]$ is the investor's recovery rate.
\ed

It is clear that $S^{i,1}_t = M^{i,1}_t$  ($S^{i,2}_t = M^{i,2}_t$, resp.) whenever $R^1_t=1$ ($R^2_t=1$, resp.), that is, when the claimed value is paid in
full by a defaulted party. We will later adjust Definition \ref{sv1} to account for the presence of a collateral, which is either posted or received
by the investor (see Definition \ref{5AAc}).

\ssc{Replacement Costs at Default}  \label{sect2.2}

Our next goal is to formalize the crucial concept of a replacement cost
of a contract in the event of counterparty's default. The following description
conveys the general idea of this notion:
the {\it replacement cost} is the cost of entering at time of counterparty's default an `equivalent'
contract enabling the non-defaulting party to keep a `similar position' in the market.
The meaning of a `similar position in the market' is ambiguous, however, and thus several alternative
specifications of the replacement cost can be postulated. Essentially, we have in mind a contract that
is as close as possible to the original one, including the counterparty risk of either one or both parties.
We argue that it is crucial to make a distinction between the convention specification of the contract's settlement value
at the time of default of a counterparty and its actual replacement cost.
For instance, the settlement clause with the original counterparty
could stipulate that in the event of default of the counterparty, if the MtM value
of the corresponding risk-free contract is positive (that is, when $M^a_t>0$), the investor receives 50\% of this value (that is, $R^2_t =.5$).
This does not mean that the replacement cost is already uniquely defined, since its computation should refer
to an `equivalent' contract   (or a `similar position') and in this example the corresponding risk-free contract
is manifestly not `equivalent' to the original contract with a defaultable counterparty, even when we agree that
the investor's own default risk could be neglected. This argument shows that,  albeit Definitions \ref{def3.6} and \ref{def3.6c} hinge on
the same collection of conventions $\{a, a', b, c, c'\}$, they have in fact completely different financial interpretations.
One can think about the first equivalent contract as the contract that is initiated when a default of one of the counterparties
in the original contract occurred and the original contract was settled.

\newpage

\bd  \lab{def3.6c}
By an {\it  equivalent contract} at time $t$, we mean a contract with identical features as the original one
in terms of promised future cash flows, but which is initiated at time $t$ between: \hfill \break
$a)$ two risk-free companies; \hfill \break
$a')$ the non-defaulting party and a risk-free company;  \hfill \break
$b)$ the non-defaulting party and a company with the same creditworthiness at time $t$ as the defaulting party
creditworthiness at the inception of the contract; \hfill \break
$c)$ the non-defaulting party and a company with the same creditworthiness at time $t$ as the defaulting party
creditworthiness just prior to default; \hfill \break
$c')$  a company with the same creditworthiness
at time $t$ as the non-defaulting party just before default and a company with the same creditworthiness at time $t$ as
the defaulting party creditworthiness just before default.
\ed

Let $P^{*,j,2}_{t}$ stand for the value of an equivalent contract, as seen by the investor,
in the event of the counterparty's default before the maturity date and prior to the investor's default,
where $j$ refers to a particular convention $a, a', b, c$ or $c'$ of Definition \ref{def3.6c} selected by the investor.
Similarly, we denote by $P^{*,k,1}_t$ the price of an equivalent contract (as seen by the investor) under convention $k)$ chosen by the counterparty,
if the investor defaults at time $t$ prior to the maturity date and before the counterparty's default.
Note that the equalities $M^{i,1}_{t}= P^{*,i,1}_{t}$ and $M^{i,2}_{t}=P^{*,i,2}_{1}$ hold, provided that the same methods (and models, if needed)
are used to compute the respective quantities. This is not likely to occur in practice, however, since the MtM value have to be calculated
using a pre-agreed common approach, whereas each party will opt for its own method to value an equivalent contract.

The ISDA Close-out Amount Protocol published in 2009 stipulates that
in determining a close-out amount, the determining party may consider any relevant information,
including, without limitation, one or more of the following types
of information (see pp. 15-16 in \cite{ISDA}): \hfill \break (i) quotations (either firm or indicative) for
replacement transactions supplied by one or more third parties that
may take into account the creditworthiness of the determining party
at the time the quotation is provided and the terms of any relevant
documentation, including credit support documentation, between the
determining party and the third party providing the quotation; \hfill \break (ii)
information consisting of relevant market data in the relevant
market supplied by one or more third parties including, without
limitation, relevant rates, prices, yields, yield curves,
volatilities, spreads, correlations or other relevant market data
in the relevant market; or \hfill \break (iii) information of the types described
in clause (i) or (ii) above from internal sources if that information is of the
same type used by the determining party in the regular course of
its business for the valuation of similar transactions.

We note that clause (i) is consistent with our view that the notion of an `equivalent contract' plays a crucial role in
the CVA computation. As argued by Parker and McGarry \cite{PG}, the close-out rules advocated by ISDA
suffer from essential weaknesses.  First, recently there have been problems in finding reference market-maker banks willing to
realistically price transactions following a major default.
Second, the `loss' method, where the determining party calculates what their loss is
using reasonableness and doing so in good faith, can bring its own problems and potential
disputes between parties as to what the loss reasonably is. Hence there is a vital need to examine
the consequences of alternative conventions that may be applied by the determining party to compute the
loss (or gain) at default.

\ssc{Loss and Gain Given Default}  \label{sect2.2v}

Loss and gain given default are aimed to quantify the effects of default on both counterparties.
Note that $i$ stands for a predetermined close-out convention adopted by both parties, whereas $j$ ($k$, resp.) denotes
a particular convention chosen by the investor (the counterparty, resp.) for internal
computations the MtM value and the value of an equivalent contract.

\bd  \lab{lgd1}
The investor's {\it loss given default} $L^{i,j,1}_t$  represents the losses incurred
by the non-defaulting investor in the event of the counterparty default at time $t$.
It is equal to the difference between the replacement cost and the
settlement value of the contract at the moment of default, that is, $L^{i,j,1}_t := P_t^{*,j,2} - S^{i,2}_t$.
The investor's {\it gain given default} $G^{i,j,1}_t$ represents the gains incurred
by the investor at time $t$ if his default occurs at this date and the counterparty has not yet defaulted.
It is given as the difference between the settlement value of the contract and the MtM value at the moment of default,
that is, $G^{i,j,1}_t := S^{i,1}_t - M^{j,1}_t$.
For the counterparty, we find it convenient to set  $L^{i,k,2}_t := S^{i,1}_t -  P_t^{*,k,1}$ and $G^{i,k,2}_t :=  M^{k,2}_t - S^{i,2}_t$,
so that all losses and gains are formally evaluated from the perspective of the investor.
\ed

It is worth stressing that we do not postulate in Definition \ref{lgd1} that the value of the loss (gain, resp.) given default
is necessarily a positive process. Also, the investor loss (gain, resp.) process and the counterparty's gain (loss, resp.) process do not mirror each other,
in general. This in turn implies that that the contract's value
is asymmetric, in general. In financial terms, this asymmetry is a natural consequence of potentially different
contingency plans in the event of default on each side of the contract.
Let $\tau^1 $ and $\tau^2$ stand for the random times
of default of the investor and the counterparty, respectively. They are defined on an underlying probability space $(\Omega , {\cal G},\Q)$,
where $\Q$ is the risk-neutral probability measure.
We denote by $\EE$ the expectation under $\Q$ and by $\G_t$ the $\sigma$-field of all events observed
by time $t$. It is assumed throughout that $\tau^1 $ and $\tau^2$ are stopping times with respect to the filtration
$\FG  = (\G_t)_{t \geq 0}$ and $\Q ( \tau^1 = \tau^2)=0$.
In order to formalize the notions of the wrong and right way risks, we need first to
formally define the loss processes associated with a contract.

\bd \lab{defwr}
The {\it loss process} $\mathcal{L}^{i,j,1}$ represents the investor's loss given default, specifically,
$\mathcal{L}^{i,j,1}_t := \I_{\{ t \ge  \tau^2 \}} \, L^{i,j,1}_{\tau^2}$.
The {\it loss process} $\mathcal{L}^{i,k,2}$ represents the counterparty's loss given default, specifically,
$\mathcal{L}^{i,k,2}_t :=  \I_{\{ t \ge \tau^1 \}} \, L^{i,k,2}_{ \tau^1 }$.
\ed

The next definition proposes a rather abstract description of the wrong-way and right-way risks.
Intuitively, the risk is {\it wrong} ({\it right}, resp.) {\it way} if the exposure tends to increase (decrease, resp.) when
the counterparty's credit quality worsens.  We will later illustrate these concepts by means of more explicit examples.

\bd \label{def2.15c}
The {\it investor's right} ({\it wrong}, resp.) {\it way risk} is the positive (negative, resp.) dependence between the counterparty's
creditworthiness  $c^2$ and the {\it positive loss} process $\mathcal{L}^{i,j,1,+}_t := \big[ \mathcal{L}^{i,j,1}_t \big]^+ $.
The {\it counterparty's right} ({\it wrong}, resp.) {\it way risk} is the positive (negative, resp.) dependence between the
investor's creditworthiness  $c^1$ and the {\it positive loss} process $\mathcal{L}^{i,k,2,+}_t := \big[ \mathcal{L}^{i,k,2}_t \big]^+ $.
\ed

\ssc{Settlement Values of Collateralized Contracts}  \label{sect2.4}

The use of collateral has become an essential risk mitigation technique
in wholesale financial markets. Financial institutions extensively employ collateral in cash
lending/borrowing activities (for instance, repo transactions), securities lending and borrowing,
and derivatives markets. In addition, central banks typically require collateral
in their credit operations (see Liberti and Mian \cite{LM}). The recent financial crisis has shed new light on
the importance of collaterals management and its impact on the overall  liquidity of the financial system.
In particular, rehypothecation was perceived to be one of the major `drivers of contagion' during the recent financial crisis.
For a comprehensive review of the liquidity crisis, we refer to Adrian and Shin \cite{AS},
Allen and Carletti \cite{AC}, Brunnermeier \cite{B}, Danielsson and Shin \cite{DS}, Ewerhart and Tapkin \cite{ET},
Gorton and Metrick \cite{GM}, and Singh and Aitken \cite{SA}.

\sssc{Uncollateralized and Collateralized Exposures}

We begin by introducing basic notions associated with the {\it Credit Support Annex} (CSA),
which is a legal document regulating credit support (that is, a {\it collateral}) for derivatives contracts,
usually in the form of a {\it margin agreement}. The goal of a CSA is to define specific clauses under which
collaterals are posted or transferred between counterparties to mitigate the exposure to the counterparty risk.

To alleviate notation, the superscript $i$ referring to the settlement convention is left out throughout Sections  \ref{sect2.4} and \ref{coven},
and thus we write $M^1_t$ and $M^2_t$, instead of $M^{i,1}_t$ and $M^{i,2}_t$.
The {\it uncollateralized exposure} for the investor (the counterparty, resp.) is thus simply
equal to $M^2_t$ ($-M^1_t$, resp.). We will consider first some simplified versions of real-world margin
agreements; for a more detailed description of the actual conventions, we refer to Section \ref{coven}.

We denote by  $C_t$ the collateral process, as seen by the investor.  Then
\bde
C_t = C^+_t - C^{-}_t = \I_{\{ C_t \geq 0\}} \, C^2_t  - \I_{\{ C_t < 0\}} \, C^1_t ,
\ede
where $C^1_t$ ($C^2_t$, resp.) denotes the market value of the basket of collaterals posted at time $t$ by the
investor (the counterparty, resp.). Note that the above decomposition is unique, meaning that the counterparties
never post collaterals simultaneously. The financial interpretation of a collateral is clear from
to the following specifications of the investor's collateralized exposures.

\bd  \lab{5AA}
The investor's {\it collateralized exposure} in the event of investor's default equals
\bde
E^{1}_t := M^1_t - C_t = \I_{\{ C_t \geq 0\}} \, (M^1_t - C^2_t) + \I_{\{ C_t < 0\}} (M^1_t + C^1_t).
\ede
The investor's {\it collateralized exposure} in the event of counterparty's default equals
\bde
E^{2}_t := M^2_t - C_t = \I_{\{ C_t \geq 0\}} \, (M^2_t - C^2_t) + \I_{\{ C_t < 0\}} ( M^2_t+ C^1_t).
\ede
\ed

Note that the collateralized exposures, which may be either positive or negative, underpin the settlement
values of a  collateralized contract, in the sense that Definition \ref{sv1} should now be applied to exposures $E^1$ or $E^2$
for some specification of the collateral process $C$.  Let us state a suitably amended  Definition \ref{sv1} of settlement values.

\bd \lab{5AAc}
In the case of investor's default at time $t$, the {\it settlement value} $S^1_t$ of a collateralized contract equals
$S^1_t := C_t + (E^1_t)^+- R^1_t (E^1_t)^-$ or, equivalently,
\beqa \lab{S11}
S^1_t &=& \I_{\{ C_t \geq 0\}} \, \Big( C^2_t + (M^1_t - C^2_t )^+ -  R^1_t \, (M^1_t - C^2_t )^- \Big)
  \\ &&\mbox{}-  \I_{\{ C_t < 0\}} \, \Big( C^1_t + R^1_t \, ( M^1_t + C^1_t )^- - (M^1_t + C^1_t)^+ \Big). \nonumber
\eeqa
In the case of counterparty's default at time $t$, the {\it settlement value} $S^2_t$ of a collateralized contract equals
$S^2_t := C_t + R^2_t (E^2_t)^+ - (E^2_t)^-$ or, equivalently,
\beqa \lab{5A}
S^2_t &=& \I_{\{ C_t \geq 0\}} \, \Big( C^2_t + R^2_t \, (M^2_t - C^2_t )^+ - (M^2_t - C^2_t )^- \Big)
 \\ &&\mbox{} -  \I_{\{ C_t < 0\}} \, \Big( C^1_t + ( M^2_t + C^1_t )^- - R^2_t \, (M^2_t + C^1_t)^+ \Big) . \nonumber
\eeqa
\ed

To simplify the presentation, we assume from now on that $M^1=M^2$ and
$\{ M_t \geq 0 \} = \{ C_t \geq 0 \} $.
In the case of a contract that is {\it fully collateralized} at time of default, we have that $C_t = M_t$ (i.e., $C^2_t = M_t^+$
and $C^1_t = M_t^-$) and thus collateralized exposures $E^1_t$ and $E^2_t$ vanish, so that
$S^1_t=S^2_t=M_t$. In the market practice, the so-called {\it under-collateralization} may occur if either the value of the collaterals decreases or the
MtM value of the contract increases, as seen by the net creditor, during the so-called {\it margin period of risk}.
In our model of a collateralized contract, we may define this concept as follows.

\bd
The investor is {\it under-collateralized} on the event $\{ E^{1}_t > 0 , M_t > 0 \}$.
The counterparty is {\it under-collateralized} on the event $\{ E^{2}_t > 0 , M_t < 0 \}$.
\ed

Collaterals mitigate risk for the receiver, but creates risk for the giver. In effect, collaterals are generally `pledged',
which means that the posting of collaterals corresponds to a `transfer of title': the ownership
of collaterals is transferred from the net debtor to the net creditor. This implies that, in the event of the net creditor's
bankruptcy, the net debtor becomes an unsecured creditor with regard the collaterals in
excess over the contract exposure. This feature of a  collateralized contract  is the source of {\it over-collateralization}.

\bd
The investor is {\it over-collateralized} on the event $\{ E^{1}_t > 0 , M_t < 0 \}$.
The counterparty is {\it over-collateralized} on the event $\{ E^{1}_t < 0 , M_t > 0 \}$.
\ed

Contracts are commonly over-collateralized because of the {\it haircut}. In effect,
the haircut, a risk mitigation tool against under-collateralization to the benefit of the net creditor,
creates over-collateralization risk for the net debtor, unless the collaterals are managed in a {\it segregated account}.

\sssc{Lock-Up Margins}

We will now complement our previous analysis by considering another potential clause of the CSA, namely,
the {\it lock-up margin}.

\bd
The {\it lock-up margin amount} $U_t$ at time $t$ is a pledged amount that is excluded from
the exposure calculation before the moment $\tau^1 \wedge \tau^2$ of the first default,
but it is accounted for in the settlement procedure in the event of the first default.
\ed

Let $U^1_t$ ($U^2_t$, resp.) be the lock-up margin posted at time $t$ by the investor (the counterparty, resp.).
We denote by $V_t = U_t^2 - U_t^1$ the net value of lock-up margins, as seen by the investor.

\bl
In the case of the investor's default at time $t$, the {\it settlement value} $S^1_t$ of a collateralized contract with
the net lock-up margin $V_t$ is
equal to
\beqa \lab{S12}
S^1_t & = &  \I_{\{ M_t  \geq  0\}} \, \Big( C^2_t  - V_t+
(M_t - C^2_t + V_t)^+   -  R^1_t \, (M_t - C^2_t + V_t)^- \Big)
\\ &&\mbox{}-  \I_{\{ M_t  < 0\}} \, \Big( C^1_t + V_t  -
(M_t + C^1_t + V_t)^+  +  R^1_t \,  (M_t + C^1_t + V_t)^- \Big) . \nonumber
\eeqa
\el

\proof
We now set $E^1_t = M_t - (C_t-V_t)$ where $C_t = \I_{\{ M_t > 0\}} \, C^2_t  - \I_{\{ M_t < 0\}} \, C^1_t$, so that
\be \lab{xde}
E^{1}_t = \I_{\{ M_t \geq 0\}} \, (M_t - C^2_t +V_t ) + \I_{\{ M_t < 0\}} (M_t + C^1_t +V_t)
\ee
and
\bde
S^1_t = C_t - V_t + (E^1_t)^+- R^1_t (E^1_t)^-.
\ede
This yields the asserted equality (\ref{S12}). Note that for $R^1_t=1$, we obtain $S^1_t = M_t$, as expected.
\endproof

\bl \lab{5D1}
In the case of the counterparty's default at time $t$, the {\it settlement value} $S^2_t$ of a collateralized contract with
the net lock-up margin $U_t$ is given by
\beqa \lab{7A}
S^2_t & = &  \I_{\{ M_t  \geq 0\}} \, \Big( C^2_t - V_t+ R^2_t \, (M_t - C^2_t + V_t)^+   -  (M_t - C^2_t + V_t)^- \Big)
\\  &&\mbox{}-   \I_{\{ M_t  < 0\}} \, \Big( C^1_t + V_t- R^2_t \, (M_t + C^1_t + V_t)^+  + (M_t + C^1_t + V_t)^- \Big).
  \nonumber
\eeqa
\el

\proof
We now combine (\ref{xde}) with the equality $S^2_t = C_t - V_t + R^2_t (E^1_t)^+-  (E^1_t)^-$.
Observe that for $R^2_t=1$, we get $S^2_t = M_t$.
\endproof

\sssc{Segregation of Collaterals}

Next, we examine the settlement values when the CSA stipulates that collaterals should be {\it segregated}.

\bd
The holding of a collateral in a risk-free account strictly separated from the net creditor's own accounts
is termed the {\it segregation} of collateral.
\ed

The goal of segregation is to mitigate the over-collateralization risk for the net debtor party, forasmuch as the third party is
assumed to be risk-free. Notice that this assumption could be also be relaxed, by accounting for the probability of default
and recovery rate of the third party custodian.

\bl
In the case of investor's default at time $t$, the {\it settlement value} $S^1_t$ of a collateralized contract with segregation
is given by the formula
\be \lab{S13}
S^1_t = \I_{\{ M_t  \geq 0\}} \, M_t -  \I_{\{ M_t  < 0\}} \, \Big( C^1_t + R^1_t \, ( M_t + C^1_t )^- - (M_t + C^1_t)^+ \Big).
\ee
\el

\proof
We first observe that segregation is only relevant for the first term in the right-hand side of (\ref{S11}).
Assume that there is over-collateralization at time $t$, then the tri-party returns part,
or the integrality, of the basket of collaterals, depending on whether the MtM
value at settlement time is positive or negative. Consequently,
the recovery rate $R^1_t$ in the first term of (\ref{S11}) equals 1, and thus equality (\ref{S11}) becomes
\beq
S^1_t &=& \I_{\{ M_t  \geq 0\}} \, \Big( C^2_t + (M_t - C^2_t )^+ -  (M_t - C^2_t )^- \Big)
\\&&\mbox{} -  \I_{\{ M_t  < 0\}} \, \Big( C^1_t + R^1_t \, ( M_t + C^1_t )^- - (M_t + C^1_t)^+ \Big).
\eeq
This establishes the result.
\endproof

\bl
In the case of counterparty's default at time $t$, the {\it settlement value} $S^2_t$ of a collateralized contract is given by
\be \lab{8A}
S^2_t = \I_{\{ M_t \geq 0\}} \, \Big( C^2_t + R^2_t \, (M_t - C^2_t )^+
- ( M_t - C^2_t )^- \Big)  +  \I_{\{ M_t < 0\}} \, M_t .
\ee
\el

\proof
Segregation cancels the risk of over-collateralization. Therefore, it suffices to set $R^2_t =1$
in the second term in the right-hand side of (\ref{5A}).
\endproof

\sssc{Lock-Up Margins and Segregation of Collaterals}

Finally, we derive the settlement values when both the lock-up margins and the segregation of collaterals
are assumed.

\bl \lab{5D2}
The {\it settlement value} $S^1_t$ of a collateralized contract with lock-up margins and segregation,
in the case of the investor's default at time $t$, is given by the following expression
\be  \lab{V9A}
S^1_t =   \I_{\{ M_t \geq 0\}} \, M_t -  \I_{\{ M_t  < 0\}}
\, \Big( C^1_t + V_t - (M_t + C^1_t + V_t)^+  + R^1_t \,  (M_t + C^1_t + V_t)^- \Big) .
\ee
\el

\proof
Equality (\ref{V9A}) is established by substituting $R^1_t$ with 1 in the first term of the right-hand side of equality (\ref{S12}).
\endproof

\bl \lab{5D3}
The {\it settlement value} $S^2_t$ of a collateralized contract with lock-up margins and segregation,
in the case of the counterparty's default at time $t$, is given by the formula
\be  \lab{9A}
S^2_t  =   \I_{\{ M_t  \geq 0\}} \, \Big( C^2_t - V_t + R^2_t \,(M_t - C^2_t + V_t)^+  - (M_t - C^2_t + V_t)^- \Big)
+ \I_{\{ M_t  < 0\}} \, M_t .
\ee
\el

\ssc{Path-dependent Features of Margin Agreements}  \label{coven}

The collateral process $C$ arising from a typical CSA appears to be a path-dependent
functional of the MtM process, rather than a function of the current MtM value of a contract. This phenomenon is due to specific features of
the actual margin agreements, which we now briefly summarize in this section.

\bd
A {\it margin agreement} is a legally binding contract that requires one or both obligors to post at time $t$,
on the event $\{t < \tau^1 \wedge \tau^2 \wedge T\}$, collaterals $C_t$ when the uncollateralized exposure $E_t$ exceeds a
predetermined {\it threshold exposure} $H_t$.
If the excess further increases, an additional collateral must be posted. By contrast,
if the excess declines then a part of the posted collateral must be returned.
The {\it minimum transfer amount} $A_t$ at time $t$ is the minimum amount under
which an additional collateral is not posted, nor a pledged collateral returned.
\ed

In market practice, at the contract's inception, collateral thresholds $H^1$ and $H^2$, as well as minimum transfer
amounts $A^1$ and $A^2$ are set for each obligor and they are based on their initial credit qualities. The more sound
credit position of a party, the higher its threshold and the minimum transfer amount.
In addition, the contract's CSA usually includes a {\it termination option} in case of a specific credit event,
such as a rating downgrade, which allows for re-negotiation of the contract's clauses.
For example, in the case of the investor's credit rating downgrade, the counterparty would
probably exercise his termination option and re-negotiate
more stringent haircut and minimum transfer amount for the investor.
As a result, collateral thresholds and  minimum transfer
amounts can be seen as stochastic processes dependent on creditworthiness processes $c^1$ and $c^2$.

\sssc{Margin Periods of Risk}

Other essential features of a margin agreement are the margin period of risk for the
collateral receiver  (i.e., the {\it net creditor}) and the margin period of risk for the collateral giver
 (i.e., the {\it net debtor}).

\bd
On the event of the collateral giver (resp. receiver) defaulting, the {\it margin period of risk for the collateral
receiver} (resp. {\it giver}) is the lag between the last time when collaterals were called for and the moment
when the collaterals are sold in the market.
\ed

The margin period of risk accounts for the frequency at which collaterals are called for (typically, one day),
plus the time necessary for accessing collaterals and appropriately selling them. In effect,
the access to the collaterals may be hindered by legal or operational obstacles, such as a
`freezing period' set by a bankruptcy court, or the segregation of the collaterals in a tri-party account.

In addition, it is customary to allow for the collaterals not to be sold immediately, as the moment of default
may happen during financial stress. In essence, the holder of collaterals is charged, for the mutual
benefit of the obligors, with the duty of selling the collaterals at a moment reasonably close to
default, which maximizes the proceeds from selling collaterals.
It is generally agreed that a period of two weeks is a good estimate for the margin period of risk
for the collateral receiver.

\sssc{Haircuts}

As was already explained, collaterals mitigate risk for the receiver (i.e., they decrease under-collateralization),
but create risk for the giver. In effect, collaterals are generally `pledged',
which means that the posting of collaterals corresponds to a `transfer of title': the ownership
of collaterals is transferred from the net debtor to the net creditor. This implies that, in the event of the net creditor's
bankruptcy, the net debtor becomes an unsecured creditor with regard the collaterals in
excess over the contract exposure. This excess is the source of over-collateralization.
Real-life contracts are commonly over-collateralized because of the so-called {\it haircut}. In effect,
the haircut, which is a risk mitigation tool against under-collateralization to the benefit of the net creditor,
creates over-collateralization risk for the net debtor, unless collaterals are managed by
a tri-party in a segregated account.

\bd
The percentage that is subtracted from the par value of the assets that are being posted as collateral,
so as to account for the perceived risk, for the collateral receiver, associated with holding
the assets during the margin period of risk, is called the {\it haircut} or the {\it valuation percentage}.
\ed

For example, the parties may decide to impose a $1\%$ haircut on government bonds and a $25\%$ haircut
on common stocks. The former type of collaterals is more common than the latter, though. In effect,
the most popular collaterals within the financial market are cash and government bonds.
First, these assets are easily valued, which facilitates the computation of the collateralized exposure.
Second, their volatility is relatively low, which diminishes both the risk of under-collateralization for
the collaterals receiver and  the risk of over-collateralization for the collaterals giver.
Finally, these assets are easily transferable, which facilitates operations.
The obligors may authorize other types of collaterals, like stocks or securitized derivatives,
which are more volatile and subject to systemic risk, but offer benefits in terms of diversification.
As a matter of fact, in theory, the breadth of available collateral types should enable mitigation
of both under- and over-collateralization risks through adequate diversification within a basket of collaterals,
which is then characterized by its {\it average haircut}.
However, as emphasized by Gorton \cite{G1,G2,G3}, an abrupt change in the level of haircuts (in particular, for repo transactions)
may lead to liquidity squeeze and, in consequence, to a systemic contagion effect.
Let us denote the average haircut processes for the investor and the counterparty by $h^1$ and $h^2$,
respectively.

\bd
The {\it average haircut} $h^j_t$ for party $j$ is the average percentage of haircut for the
basket of posted collaterals at time $t$.
\ed

It is frequently assumed in the literature (see, for instance, Pykhtin and Zhu \cite{PZ})
 that the variation of the value of the basket of collaterals is negligible
during the margin period of risk. This assumption, which offers benefits in terms of simplification (especially for simulations),
is equivalent to the statement that the variation of the
collateralized exposure during the margin period of risk is approximately equal to the
variation of the contract's MtM value. If, during the margin period of risk, the variation of the collateralized exposure is approximately
equal to the variation of the contract's MtM value then the collateral process for
the investor can be approximated as follows (note that the formulae below differ slightly from results of Pykhtin and Zhu \cite{PZ})
\begin{equation*}
C^1_t \simeq  \left\{ \begin{array}{ll} \big( E_t^2-(H^1_t+A^1_t)\big)^+  & \mbox{ for }\ t \in [0, \tau^1 \wedge \tau^2 \wedge T],\\
(1+h^1_{\tau^1 \wedge \tau^2 }) \, \big( E^2_{\tau^1 \wedge \tau^2 } -(H^1_{\tau^1 \wedge \tau^2}+A^1_{\tau^1 \wedge \tau^2})\big)^+&
\mbox{ for }\ t \in \, ]\tau^1 \wedge \tau^2  ,
\tau^1 \wedge \tau^2+s ], \end{array} \right.
\end{equation*}
where $s$ denotes the common value of the  margin period of risk and $h^1_t$ ($h^2_t$, resp.) is the average basket haircut
for the investor (counterparty, resp.) at time $t$. By the same token, the collateral process for the counterparty satisfies
\begin{equation*}
C^2_t \simeq  \left\{ \begin{array}{ll} \big( E_t^1-(H^2_t+A^2_t)\big)^+  & \mbox{ for }\ t \in [0, \tau^1 \wedge \tau^2 \wedge T],\\
(1+h^2_{\tau^1 \wedge \tau^2 }) \, \big( E^1_{\tau^1 \wedge \tau^2 } -(H^2_{\tau^1 \wedge \tau^2}+A^2_{\tau^1 \wedge \tau^2})\big)^+& \mbox{ for }\ t \in \, ]\tau^1 \wedge \tau^2  ,
\tau^1 \wedge \tau^2+s ]. \end{array} \right.
\end{equation*}
The assumptions stated above are supported by the fact that the volatility of the basket of collaterals
is generally lower than the volatility of the underlying contract value. In effect, first,
a basket of collaterals usually comprises a significant amount of cash and government bonds,
which are generally assets with low volatility and, second, the exposure to the basket of collaterals
can in theory be mitigated through diversification, contrary to the exposure to the MtM
value of the contract. Let us consider, for instance, a basket of collaterals made of long positions in stock indices
and government bonds. Such a basket would have offered useful diversification features when Lehman
Brothers went bankrupt, as the investors, by `going safe', made the values of stocks decrease and the
prices of the government bonds increase. It would not have necessarily offered
similar diversification features during the recent Greek sovereign debt crisis, however, when both stock indices and government
bonds had the tendency to decrease concurrently. These stylized examples show that a suitably diversified basket of
collaterals requires choices tailored to both the contract at hand and the counterparty. In particular, a
great care must be taken for avoiding wrong way risk events in regard of collaterals.

Let us conclude this section by making a few more conjectures on the behavior of collaterals.
If the under-collateralization risk is suitably diversified within
the basket of collaterals, the expectation of the downside variation
in collaterals value, in the event of default of the net debtor
and during the margin period of risk for the net creditor, can be set to zero.
Similarly, if the over-collateralization risk is suitably diversified within the
basket of collaterals, the expectation of the upside variation in collaterals value,
in the event of default of the net creditor and during the margin period of risk for the net debtor, can be set to zero.
In view of these observations, we formulate a plausible conjecture, which was also established
in a slightly different version by Pykhtin and Zhu \cite{PZ}.
If both under-collateralization and over-collateralization risks are suitably
diversified within the baskets of collaterals, then the collateral process for
the investor can be approximated as follows
\begin{equation*}
C^1_t \simeq  \left\{ \begin{array}{ll} (1+h^1_t) \,\big( E_t^2-(H^1_t+A^1_t)\big)^+  & \mbox{ for }\ t \in [0, \tau^1 \wedge \tau^2 \wedge T],\\
(1+h^1_{\tau^1 \wedge \tau^2 }) \, \big( E^2_{\tau^1 \wedge \tau^2 } -(H^1_{\tau^1 \wedge \tau^2}
+A^1_{\tau^1 \wedge \tau^2})\big)^+& \mbox{ for }\ t \in \, ]\tau^1 \wedge \tau^2  ,
\tau^1 \wedge \tau^2+s ], \end{array} \right.
\end{equation*}
whereas for the collateral process for the counterparty we obtain
\begin{equation*}
C^2_t \simeq  \left\{ \begin{array}{ll} (1+h^2_t) \, \big( E_t^1-(H^2_t+A^2_t)\big)^+  & \mbox{ for }\ t \in [0, \tau^1 \wedge \tau^2 \wedge T],\\
(1+h^2_{\tau^1 \wedge \tau^2 }) \, \big( E^1_{\tau^1 \wedge \tau^2 } -(H^2_{\tau^1 \wedge \tau^2}+A^2_{\tau^1 \wedge \tau^2})\big)^+& \mbox{ for }\ t \in \, ]\tau^1 \wedge \tau^2  ,
\tau^1 \wedge \tau^2+s ]. \end{array} \right.
\end{equation*}

\ssc{Symmetric Case} \lab{spec}

To reduce the variety of cases under study, we will frequently postulate in numerical implementations that $i=j=k$,
meaning that the settlement values and the replacement costs refer to the same convention $i)$.
Suppose that we assume, in addition, that $M^{i} = P^{*,i,1}= P^{*,i,2}$, so that MtM value and replacement costs coincide.
It is worth stressing that these equalities can be identified as the tacit assumptions made in papers that omit the replacement cost,
only refer to the MtM value upon default and consider a unique convention for the MtM value.
Under these simplifying assumptions, the equalities $L^{i,1}:= P^{*,i,2}- S^{i,2}= M^{i,2}-S^{i,2} =: G^{i,2}$, as well as $L^{i,2}= G^{i,1}$, are manifestly satisfied and thus the pricing problem becomes symmetric.
Once again, to alleviate notation we leave out the superscript $i$ and we first consider an  uncollateralized contract.

\bl
On the event  $ \{ \tau^2 = t \le T, \tau^1 > \tau^2 \}$ of counterparty's default, the
loss given default for the investor equals $L^{1}_t = (1-R^2_t) \, M_t^+$
where $R^2_t$ is the counterparty's recovery rate at time $t$.
On the event  $ \{ \tau^1 = t \le T, \tau^2 > \tau^1 \}$ of investor's default, the gain given default
for the investor equals $G^1_t = (1-R_t^1) \, M_t^-$
where $R^1_t$ is the investor's recovery rate at time $t$.
\el

\proof
We substitute the equivalent contract price to the MtM value in Definition  \ref{sv1}
of the settlement value and we represent the settlement value in terms of the equivalent contract price,
so that
\be
S^2_t = R_t^2 \, M_t^+ - M_t^- = M_t - (1-R_t^2) \, M_t^+. \lab{sv01}
\ee
From Definition \ref{lgd1}, we obtain $L^{1}_t  = M_t  - S^2_t$ and thus the stated formula follows.
Similarly,
\be
S^1_t = M_t^+ - R_t^1 \, M_t^-  = M_t + (1-R_t^1) \, M_t^- ,  \lab{sv02}
\ee
and the asserted formula now follows since $G^{1}_t = S^1_t - M_t$.
\endproof

In view of Definition \ref{lgd1}, for the counterparty we obtain
$L^{2}_t = (1-R^1_t) \, M_t^-$ and $G^2_t = (1-R_t^2) \, M_t^- $.
The next result provides the loss and gain given default, as seen by the investor, in the presence of a collateral agreement.

\bl
On the event  $ \{ \tau^2 = t \le T, \tau^1 > \tau^2 \} $ of counterparty's default,
the loss given default for the investor equals
\beq 
L^1_t = (1- R^2_t ) \, \Big(  \I_{\{ M_t \geq 0\}} \, (M_t - C^2_t )^+ + \I_{\{ M_t  < 0\}} \, (M_t + C^1_t )^+\Big).
\eeq
On the event  $ \{ \tau^1 =t  \le T, \tau^2 > \tau^1 \}$ of investor's default, the gain given default
for the investor equals
\beq \lab{B}
G^1_t = (1 - R^1_t ) \, \Big( \I_{\{ M_t \geq 0\}} \, ( M_t - C^2_t )^-
 + \I_{\{  M_t < 0\}} \, ( M_t + C^1_t ) ^- \Big).
\eeq
\el

\proof
The lemma follows by a direct application of formulae (\ref{S11}) and (\ref{5A}) to Definition \ref{lgd1}.
\endproof

The next result holds under segregation of collaterals.

\bl
On the event $ \{ \tau^2 =t \le T, \tau^1 > \tau^2 \} $ of the counterparty's default, the loss given default for the investor equals
\bde
L^1_t = \I_{\{ M_t \ge 0\}} \, (1-R^2_t ) \, (M_t - C^2_t)^+ .
\ede
\el

\proof
By straightforward computations, we obtain
\beq
L^1_t &=&M_t - S^2_t \\ &=& M_t -   \I_{\{ M_t \geq 0\}} \, \Big( C^2_t + R^2_t \,
(M_t - C^2_t)^+ - (M_t - C^2_t )^- \Big)
-  \I_{\{ M_t  < 0\}} \, M_t
\\ &=& M_t - \I_{\{ M_t \geq 0\}} \, \Big( C^2_t + R^2_t \,
(M_t - C^2_t )^+ - (M_t - C^2_t)^- \Big)
+ M_t^-
\\ &=& M_t^+ - \I_{\{ M_t\geq 0\}} \, \Big( C^2_t + R^2_t
\, (M_t - C^2_t )^+ - (M_t - C^2_t )^- \Big)
\\ &=& \I_{\{ M_t \geq 0\}} \Big( M_t -  C^2_t - R^2_t
\, (M_t - C^2_t )^+ + (M_t- C^2_t )^-  \Big)
\\ &=& \I_{\{ M_t \geq 0\}} \, (1 - R^2_t ) \, ( M_t - C^2_t )^+
 \eeq
as was asserted.
\endproof

\ssc{CVA for Bilateral Counterparty Risk}  \label{sect2.6}

Generally speaking, the {\it Credit Value Adjustment} (CVA) is the correction that should be made to the risk-free price, that is, to the price
of the contract with no counterparty risk, in order to account for the potential losses of the parties at the moment of default,
be the default triggered by the investor or the counterparty. Although the CVA computations can be performed for any choice
of conventions, to make pricing formulae more transparent, we postulate hereafter that $i=j=k$. It will still be the case that the
choice of a particular convention will have non-trivial implications on the value of the CVA, which under the present assumptions is defined as follows.

\bd \lab{cvai}
Let $P_t$ denote the price at time $t$ of a contract with no counterparty risk and let $\bhP_t$ stand for the price of
the corresponding contract with bilateral counterparty risk. Then the {\it credit value adjustment}
at time $t$ is given by the equality $\CVA_t = P_t - \bhP_t$.
 More specifically, we denote by $\CVA^i_t$ and $\bhP^i_t$, respectively, the CVA and the price of a contract with bilateral
counterparty risk for the convention $i =a, a', b, c, c'$, that is,
$\CVA ^i_t := P_t - \bhP^i_t$.
\ed

Let us stress that we depart from the usual claim in the existing literature (see, for instance, Brigo and Capponi \cite{BC})
that the CVA under the uncollateralized full two-way payment rule
equals $P_t(\mathcal{L}^{1,+})- P_t(\mathcal{L}^{2,+})$, where $P_t(\mathcal{L}^{1,+})$ ($P_t(\mathcal{L}^{2,+})$, resp.) is
the price of the protection leg of a contingent CDS  on the investor's (counterparty's, resp.) positive loss.

To proceed, we need to compute the price $\bhP^i_t$. To this end,
we denote by $\Pi(t,T)$ the {\it discounted cash flows}  over the time period $[t,T]$ of the equivalent risk-free contract,
as seen by the investor. Note that the cash flows occurring at times $t$ and $T$ are included in  $\Pi(t,T)$.
However, for any time $u$ after the current date $t$ and before the contract's maturity $T$ (that is, for any $u$ such that $t < u < T$),
by a slight abuse of notation, we denote by $\Pi(t,u)$ the cash flows that occur over the time period $[t,u)$, that is, with the date $u$ excluded.
By the same convention, in the discounted cash flows $\Pi(t,\tau_1 )$ and  $\Pi(t,\tau_1 )$  the payoffs falling
at random times $\tau_1$ and $\tau_2$ are excluded.
We first define an auxiliary concept of the bilateral first-default-free contract.
As usual, we take the position of the investor.

\bd \label{defcc}
The {\it bilateral first-default-free contract} under convention $i$) is the contract in which the investor (resp. the counterparty)
is protected against a loss due to the counterparty's default (resp. the investor's default) but not against the potential
 losses of the equivalent contract under convention $i$). Therefore, the cash flows of the contract are
\beq
\tilde{\Pi}^i (t,T) &:=& \I_{\{ \tau^1 \wedge \tau^2 > T \}} \, \Pi(t,T) + \I_{\{ \tau^1 \le T, \tau^2 > \tau^1\}} \bigg(
 \Pi(t,\tau^1) + \frac{B_t}{B_{\tau^1}} P^{*,i,1}_{\tau^1} \bigg)
 \\ && \mbox{} + \I_{\{ \tau^2 \le T, \tau^1 > \tau^2 \}}  \bigg( \Pi (t, \tau^2)
  + \frac{B_t}{B_{\tau^2}} P^{*,i,2}_{\tau^2} \bigg)
\eeq
where  $P^{*,i,2}$ and $P^{*,i,1}$ are the replacement costs under convention $i$).
The value at time $t$ of the bilateral first-default-free contract under convention $i$) is denoted by $\tilde{P}^i_t$.
 \ed

Note that for  $i=a$ we obtain $\tilde{P}^i_t = {P}_t$, as expected.
The next result furnishes a convenient generic representation for the CVA of an
uncollateralized or a collateralized contract with bilateral counterparty risk.

\bp \lab{PC9}
The price $\bhP^{i}_t$ of a contract with bilateral counterparty risk under convention i) is equal to the difference
of the price $\tilde{P}^{i}_t$ of the first-default-free contract under convention i) and the price of the protection leg of a contingent CDS
written on the investor's loss $\mathcal{L}^{i,1}$ due to the counterparty's default before the contract termination $T$,
augmented by the price of the protection leg of a contingent CDS
written on the counterparty's loss $\mathcal{L}^{i,2}$ due to  the investor's default before the contract termination $T$.
This means that the following equality holds, for every $t \in [0,T]$ on the event $\{ t < \tau^1 \wedge \tau^2 \}$,
\be \lab{uuu}
\bhP^i_t = \tilde{P}^i_t + P_t(\mathcal{L}^{i,2}) - P_t(\mathcal{L}^{i,1}).
\ee
Consequently, the credit value adjustment for a contract with bilateral counterparty risk under convention i) is given by
\be \lab{PB1}
\CVA^i_t = P_t - \tilde{P}^i_t  + P_t(\mathcal{L}^{i,1}) - P_t(\mathcal{L}^{i,2}).
\ee
\ep

\proof In view of the interpretation of the settlement values $S^{i,1}_t$ and $S^{i,2}_t$, we obtain
\beq
\bhP^i_t&=& \EE \bigg[  \I_{\{ \tau^1 \wedge \tau^2 > T \}} \, \Pi(t,T) +  \I_{\{ \tau^1 \le T, \tau^2 > \tau^1\}}
 \, \bigg( \Pi(t,\tau^1) + \frac{B_t}{B_{\tau^1}} \, S^{i,1}_{\tau^1} \bigg)
\\ && \mbox{} +   \I_{\{ \tau^2 \le T, \tau^1 > \tau^2 \}} \, \bigg( \Pi(t,\tau^2) + \frac{B_t}{B_{\tau^2}} \, S^{i,2}_{\tau^2} \bigg) \,\Big|\, \G_t \bigg]
\\ &=&  \EE \bigg[ \I_{\{ \tau^1 \wedge \tau^2 > T \}} \, \Pi(t,T) +  \I_{\{ \tau^1 \le T, \tau^2 > \tau^1\}}
 \, \bigg( \Pi(t,\tau^1) + \frac{B_t}{B_{\tau^1}} \, \big( P_{\tau_1}^{i,*,1}+L^{i,2}_{\tau_1} \big) \bigg)
\\ && \mbox{} +  \I_{\{ \tau^2 \le T, \tau^1 > \tau^2 \}} \, \bigg( \Pi(t,\tau^2) + \frac{B_t}{B_{\tau^2}}  \, \big(
P_{\tau_2}^{i,*,2} - L^{i,1}_{\tau_2} \big) \bigg)  \,\Big|\, \G_t \bigg]
\\ &=& \tilde{P}^i_t + P_t(\mathcal{L}^{i,2}) - P_t(\mathcal{L}^{i,1})
\eeq
where we used the equalities $ S^{i,1}_t = P_t^{i,*,1}+L^{i,2}_t$ and $S^{i,2}_t = P_t^{i,*,2} - L^{i,1}_t$ (see Definition \ref{lgd1}).
Hence the asserted equality (\ref{uuu}) follows. \endproof

Let us comment briefly on equality (\ref{PB1}), by focusing on the unilateral case. If one applies the formula $\CVA^i_t = P_t(\mathcal{L}^{i,1})$
to compute the CVA, one implicitly assumes that the equality $P_t = \tilde{P}^{i}_t$ holds,
which is not a viable assumption under convention $i)$, in general. Indeed, with the exception of convention $a)$,
the price of the contract with counterparty risk under convention $i)$
is a function of the price of the equivalent contract under convention $i)$, which itself is a function
of an `equivalent' contract under the same convention, and so on. Put another way, the loss at time of the counterparty's default
is a function of the potential loss of the subsequent equivalent contract, which itself is a function of the loss of the `equivalent' contract for the
first `equivalent' contract, and so on. This convolution property of pricing under convention $i)$ gives rise to a serious challenge,
and it explains why it is by far more common to adopt a simplified convention $a)$, which assumes that the equivalent contract is risk-free. Indeed,  under convention $a)$ the convolution property
disappears and thus one can identify the CVA with the value of the protection leg of the CDS contingent claim
written on the loss process.

%

\ssc{CVA Computation under Rank $n$ Pricing Rule} \lab{rankn}

Practical arguments justify the assumption that only a limited number of potential `equivalent' counterparties exist for a given contract.
Also, the required simulations under convention $a'), b), c)$ and $c')$ make it necessary to take a view on the value
of the $n$th equivalent contract for some $n \ge 1$, where by the $n$th equivalent contract we mean an `equivalent' contract
used to replace the $(n-1)$th `equivalent' contract after the default of the $(n-1)$th counterparty.
We thus find it natural and convenient to introduce the notion of the {\it rank $n$} pricing rule.
Note that this rule can be seen as either a computational tool that facilitates implementations
or a realistic feature of the market model, which reflects the fact
that the number of potential equivalent counterparties is limited.

\bd \lab{deftt}
The price of a contract with bilateral counterparty risk under the {\it rank} $n$ pricing rule is equal to
the price computed under the assumption that the $n$th equivalent contract is either given by some
predetermined value or by the value of a risk-free contract.
\ed

Of course, the definition above is merely aimed to convey the general idea of the rank $n$ pricing rule.
For explicit computations, we can use one of the following specifications, whichever is found to be practically appealing
and/or suitable for an efficient numerical procedure. The following two alternative specification will be used in
what follows.  In the foregoing definitions, we fix a settlement convention $i)$ and a natural number $ n \ge 1$.

\bd
The price of a contract under the {\it risk-free rank $n$} pricing rule
is equal to the price that is derived under the assumption that the $n$th equivalent contract equals the risk-free contract.
\ed

\bd
The price of a contract under the {\it zero rank $n$} pricing rule
is equal to the price that is derived under the assumption that the $n$th equivalent contract has value zero.
\ed

The following auxiliary result is an immediate consequence of Definition \ref{deftt} and thus its proof
is omitted. It is worth recalling that the settlement values
$S^1_t$ and $S^2_t$ are functions of the mark-to-market value $M^i_t$ and recovery processes $R^1_t$ and $R^2_t$.
To emphasize the dependence on $M^i_t$, we write $S^1_t = S^1_t  (M^i_t)$ and $S^2_t = S^2_t(M^i_t)$.

\bp \lab{44t}
The price of a contract with the bilateral counterparty risk under the settlement convention i)
and under the rank $n$ pricing rule equals $\bhP^{i,n}_t = \EE \big[ \Pi^{i,n}(t,T) \mid \G_t \big]$, for any $n \ge 1$ and $i=a,a',b,c,c'$,
where the discounted cash flows $\Pi^{i,n}(t,T)$ are given by
\beq
\Pi^{i,n}(t,T) &:=& \I_{\{ \tau^1 \wedge \tau^2 > T \}} \, \Pi(t,T)
+  \I_{\{ \tau^1 \le T, \tau^2 > \tau^1\}} \bigg[ \Pi(t,\tau^1) + \frac{B_t}{B_{\tau^1}} \, S^1_{\tau^1}\big( M^{i,n}_{\tau^1} \big) \bigg] \\
&&\mbox{}+ \I_{\{ \tau^2 \le T, \tau^1 > \tau^2 \}} \, \bigg[ \Pi(t,\tau^2)
+ \frac{B_t}{B_{\tau^2}} \, S^2_{\tau^2}\big( M^{i,n}_{\tau^2} \big) \bigg]
\eeq
where in turn the quantities $M^{i,n}_{\tau^2}$ and $M^{i,n}_{\tau^1}$ are computed under the assumption that
at most $n$ equivalent contracts can default before maturity $T$ and the value of the $n$th equivalent
contract, if it exists, is either a predefined value or the value of the equivalent risk-free contract.
\ep

It is trivial to observe that the price of a contract with bilateral counterparty risk under any settlement convention $i)$ for $i =b, c, c'$,
and under risk-free rank 1 pricing rule is simply equal to the price of the contract under convention $a)$.
Although Proposition \lab{PC9} yields an illuminating representation for $\CVA^i_t$, for the purpose of
practical computations of this quantity, we will employ the method outlined in Proposition \ref{44t} combined with the natural approximation
$\CVA^i_t \simeq P_t - \bhP^{i,n}_t$ stemming from the observation that $\bhP^{i}_t \simeq \bhP^{i,n}_t$, at least for a sufficiently large $n$.

\section{CVA Computation in a Markovian Contagion Model} \lab{sect3}

In this section, we develop a general procedure for the computation of the credit value adjustment under systemic risk.
For the sake concreteness, we will later specify our study  to the case of a credit default swap and we will postulate
that the credit qualities of the investor, the counterparty, and the reference entity are identical and depend on the level of systemic risk.
It should be made clear that the generalization of the approach to the situation where the credit
qualities are not equal is rather straightforward, since it suffices to modify the pricing engine accordingly.
The contract under consideration may be seen, ahead of the 2007 credit crunch, as a stylized version of a credit default swap between Goldman
Sachs, the buyer of protection on Lehman Brothers, and AIG,  the protection seller. The possibility of a
systemic crisis is accounted for in the pricing, by considering not only the systemic impact of the potential
defaults of the latter companies, but also of other systemic companies, such as, for instance, Merrill Lynch,
Bear Stearns, or Northern Rock. Consequently, this model prices the possibility of a scenario similar
to the wrong way risk situation that was experienced by Goldman Sachs during the financial crisis, namely, the
bankruptcy of AIG at the time when Lehman Brothers, following on the path of major financial institutions
such as Bear Stearns, was on the brink of default. Recall that, as it is much documented today,
AIG stopped meeting its contractual obligations when, as the creditworthiness of Lehman Brothers kept on deteriorating,
AIG could not meet the margin calls on its credit default swaps, on Lehman Brothers, especially. Obviously, AIG was
ill prepared to a manage an abrupt increase in liquidity requirements.

\ssc{Contagion Effects}  \label{sect2.5}

Contagion effects, that is, changes in the likelihood of default of surviving names,
 can be studied either at the level of individual credit names or for large portfolios.
We first introduce the concept of the at-first-default contagion effect.

\bd
If the default of a party tends to increase (resp. decrease) the probability of default of the other, we say that there is
a {\it negative} (resp. {\it positive}) {\it at-first-default contagion effect}.
\ed


Note that this is precisely the at-first-default contagion effect, which justifies the distinction
between the settlement conventions $c)$ and $c')$ in Definition \ref{def3.6c}.


As a consequence of the recent financial crisis, the systemic risk, which is usually interpreted as the risk of cascading defaults
following the bankruptcy of a major player within the economical system, has been
given increased scrutiny among scholars and practitioners alike.
Some studies focus on details of the mechanism which has unfolded
(see Adrian and Shin \cite{AS}, Allen and Carletti \cite{AC}, Brunnermeier \cite{B}, Danielsson and Shin \cite{DS},
Ewerhart and Tapking \cite{ET}, Gorton and Metrick \cite{GM}, and Singh and Aitken \cite{SA}). Other papers shed
some light on how to model complex relationships among players (see, for instance,
Cont and Minca \cite{CM1}, Cont and Moussa \cite{CM2}, and Duffie and Zhu \cite{DZ}).

\bd
The {\it systemic contagion risk} is the risk of a `collapse' of the entire market (or its sector)
resulting from the counterparty risk contagion.
\ed

The systemic contagion effect is frequently modeled by means of a Markov chain, which
represents the successive defaults of any {\it systemic company} within a {\it systemic basket}
of companies whose defaults increase the systemic risk.
Typically, these systemic companies may be considered as leading financial institutions such
as banks or trading exchanges. One may also consider, in lieu of a systemic company's default,
any event with systemic effects, such as a disruption within the repo market, a state credit event, and so on.
As in papers by Jarrow and Yu \cite{JY} and Laurent et al. \cite{LCF}, we find it
natural and convenient to tie the credit qualities at time $t$ of surviving obligors to the number of defaults
$N_t$ of the reference entities of the systemic basket.
We underline this dependency by amending our notation: the credit qualities of the counterparties
in the presence of systemic risk will be denoted by $c^p_t(N_t)$ for $p =1,2$.

\ssc{Time-Homogeneous Markovian Contagion Model} \label{sect3.2}

For a numerically efficient and practically appealing CVA computation, we need to propose a suitable stochastic framework and
to develop numerical schemes adapted to the valuation problem at hand. Although the underpinning stochastic model will be formulated
in the continuous-time set-up and thus the general analytic formula for the CVA will involve integrals over time period $[t,T]$
(see Proposition \ref{SF11}), for the purpose of its implementation, we will also need to derive a discretized
version of the general formula amenable to numerical tests (see Proposition \ref{SF1}).

We postulate that the joint dynamics of defaults for a
 {\it systemic portfolio} of $m$ firms, the investor, the counterparty,
and the reference entity are represented by a continuous-time Markov chain $\MM = (\MM_t)_{t \in [0,T]}$
(not to be confused with the MtM value $M_t$)
with the state space
\beq
 E = \{ (j_0,0,0,0),(j_0,1,0,0),(j_0,0,1,0),(j_0,0,0,1), (j_0,1,1,0), \\ (j_0,1,0,1),(j_0,0,1,1),(j_0,1,1,1),\, j_0 =0,1,\dots , m \}.
\eeq
A generic state $(j_0, j_1,j_2, j_3)$ of the Markov chain $\MM$ has the following interpretation: \hfill \break
(a) $j_0$ is the number of defaults in systemic portfolio, \hfill \break
(b) $j_1$ is the default indicator for the investor, \hfill \break
(c) $j_2$ is the default indicator for the counterparty, \hfill \break
(d) $j_3$ is the default indicator for the reference entity.

Suppose that we have specified the intensities of transitions between states and thus we are given
the matrix of transition intensities, denoted as $\AMC_{\MM}$. We are then in a position to formally define
the continuous-time Markov process $M$ with the state space $E$ and the infinitesimal generator $\AMC_{\MM }$.
Note that $\MM$ admits the following representation $\MM_t = ( N_t , \widehat  H^1_t, \widehat  H^2_t, \widehat  H^3_t)$,
where the process $N$ represents the number of systemic defaults and takes values in $\{0,1,\dots,m\}$,
whereas the {\it default indicator} processes $\widehat H^j,\, j=1,2,3$
take values in $\{0,1\}$. We will sometimes assume that $\MM_0=(0,0,0,0)$, although, obviously, the choice of an
initial value for the Markov chain $\MM$ is arbitrary.

Next, we specify the default times by setting $\tau^i = \inf \, \{ t \in \rr_+ : \, \widehat  H^i_t = 1 \}$ for $i=1,2,3$.
Let us mention that $\tau^3 = \tau $ will play the role of default time of the reference entity (for instance, the reference
credit name in a credit default swap which we will study in what follows).
Finally, we define the {\it default counting process} $N^*_t = N_t + \widehat H^1_t + \widehat  H^2_t + \widehat  H^3_t$.
It is clear that $N^*_t$ represents the total number of defaults in the model up to time $t$ and a generic value $l$
of the process $N^*$ satisfies $l = j_0+j_1+j_2+j_3$.

Let us list the desired properties of the model. We stress that all intensities
are assumed to be specified under the risk-neutral probability measure $\Q$. \vskip 1 pt

\noindent {\bf (M.1)} Following Laurent et al. \cite{LCF}, we postulate that the default intensity
of a systemic company is the same for every entity within the systemic basket
and it is a function of time and the total number of defaults
(including the defaults of the investor, the counterparty, and the reference entity). \vskip 1 pt
\noindent {\bf (M.2)} The systemic portfolio is assumed to be time-homogeneous and the reference name can be seen as
an additional name in the systemic portfolio. The reference entity is formally isolated, however, since
we need the exact knowledge of the default time of this name. By contrast, for the original systemic portfolio of $m$
entities, it is sufficient for our purposes to know the number of defaults that occurred by time $t$ in this portfolio;
 the identities of defaulting entities are not relevant. \vskip 1 pt
\noindent {\bf (M.3)} The default intensities of the investor and the counterparty depend on
time parameter $t$ and the total number of defaults (including the default of the investor and the counterparty).\vskip 1 pt

In order to specify the generator of $\MM$, we introduce the following auxiliary quantities: \hfill \break
(i) $\gamma (t,j_0)$ representing the intensity of a new default in the systemic portfolio of $m+1$ entities
(including the reference entity), given that the total number of defaults that occurred by time $t$ equals $j_0$
(that is, on the event $\{N^*_t =j_0\}$), \hfill \break
(ii) $\widehat \gamma_1 (t,j_0)$  (resp. $\widehat \gamma_2(t,j_0)$),
which represents the intensity of default of the investor  (resp. the counterparty)
given that the total number of defaults that occurred by time $t$ equals $j_0$
(that is, on the event $\{N^*_t =j_0\}$).

\bd \lab{gtgy}
We say that transition intensities define the {\it time-inhomogeneous Markovian contagion model} if they
satisfy, for every $j_0=0, \dots ,m$,
\beq
&&\lambda ((j_0,0,0,0), (j_0+1,0,0,0)) =  a_{j_0} \gamma (t,j_0), \\
&&\lambda ((j_0,0,0,0), (j_0,1,0,0)) =   \widehat \gamma_1 (t,j_0), \\
&&\lambda ((j_0,0,0,0), (j_0,0,1,0)) =   \widehat \gamma_2 (t,j_0), \\
&&\lambda ((j_0,0,0,0), (j_0,0,0,1)) =    b_{j_0} \gamma (t,j_0), \\
&&\lambda ((j_0,1,0,0), (j_0+1,1,0,0)) = a_{j_0} \gamma (t,j_0+1), \\
&&\lambda ((j_0,1,0,0), (j_0,1,1,0)) =   \widehat \gamma_2 (t,j_0+1), \\
&&\lambda ((j_0,1,0,0), (j_0,1,0,1)) =   b_{j_0} \gamma (t,j_0+1), \\
&&\lambda ((j_0,0,1,0), (j_0+1,0,1,0)) =  a_{j_0}  \gamma (t,j_0+1), \\
&&\lambda ((j_0,0,1,0), (j_0,1,1,0)) =   \widehat \gamma_1 (t,j_0+1), \\
&&\lambda ((j_0,0,1,0), (j_0,0,1,1)) =    b_{j_0} \gamma (t,j_0+1), \\
&&\lambda ((j_0,0,0,1), (j_0+1,0,0,1)) = \gamma (t,j_0+1), \\
&&\lambda ((j_0,0,0,1), (j_0,1,0,1)) =   \widehat \gamma_1 (t,j_0+1), \\
&&\lambda ((j_0,0,0,1), (j_0,0,1,1)) =   \widehat \gamma_2 (t,j_0+1),
\eeq
\beq
&&\lambda ((j_0,1,1,0), (j_0+1,1,1,0)) =  a_{j_0} \gamma (t,j_0+2), \\
&&\lambda ((j_0,1,1,0), (j_0,1,1,1)) =    b_{j_0} \gamma (t,j_0+2), \\
&&\lambda ((j_0,1,0,1), (j_0+1,1,0,1)) =  \gamma (t,j_0+2), \\
&&\lambda ((j_0,1,0,1), (j_0,1,1,1)) =    \widehat \gamma_2 (t,j_0+2), \\
&&\lambda ((j_0,0,1,1), (j_0+1,0,1,1)) =  \gamma (t,j_0+2), \\
&&\lambda ((j_0,0,1,1), (j_0,1,1,1)) =    \widehat \gamma_1 (t,j_0+2), \\
&&\lambda ((j_0,1,1,1), (j_0+1,1,1,1)) =  \gamma (t,j_0+3),
\eeq
where
\bde
 a_{j_0} = \frac{m-j_0}{m+1-j_0}, \quad b_{j_0} = \frac{1}{m+1-j_0}.
\ede
By the {\it time-inhomogeneous Markovian contagion model} we mean the Markov chain $\MM$ on the state space $E$ with
the generator $\AMC_{\MM }$ specified by the transition intensities given above.
\ed

In what follows, the filtration $\FG$ is assumed to be the natural filtration of the Markov chain $\MM$.
Let us note that for the Markov chain $\MM$ satisfying Definition \ref{gtgy}, the process $\bar N_t = N_t + \widehat H^3_t$
does not follow a Markov chain, in general, unless the equalities $\widehat \gamma_1 (t,j_0) = \widehat \gamma_2 (t,j_0) =0$
hold. By the same token, the default counting process $N^*$ is not necessarily a Markov chain.
We have, however, the following result, which corresponds to the case
when the investor and the counterparty can be seen as belonging to the systemic portfolio.
Observe, in particular, that equality (\ref{gtg}) means that the intensity of a new default in the original systemic
portfolio (including the reference name) is proportional to the number of surviving names
in this portfolio. Note also that a number $l \in \{0,1,\dots , m+3 \}$ represents the total number of defaults, that is,
a generic value of the process $N^*$. Recall that for a generic state $(j_0, j_1,j_2, j_3)$ of the Markov chain $\MM$
we have that $l =  j_0+j_1+j_2+j_3$.

\bl \lab{lmm}
(i) Let $ \widehat \gamma (t,l),\, t \in [0,T],\, l=0,1,\dots ,m+2$ be a strictly positive function.
Assume that $\widehat \gamma_1 (t,l) = \widehat \gamma_2 (t,l) = \widehat \gamma (t,l)$
and  the equality
\be \lab{gtg}
\lambda ((j_0, j_1,j_2,1),(j_0+1, j_1,j_2,1))  =(m-j_0) \widehat \gamma (t,l)
\ee
holds for $j_0=0,1,\dots, m-1$ and $j_1,j_2 = 0,1$, where $l = j_0+j_1+j_2+1$.  Then the default counting process $N^*$ is a time-inhomogeneous
Markov chain with state space $E^* = \{0,1,\dots , m+3 \}$ and the following
generator
\beq
\AMC_{N^*} = \left[
\begin{matrix}
- \lambda^*(t,0)  & \lambda^*(t,0) &  0  &  \dots    & 0  & 0  \\
0 & - \lambda^*(t,1) & \lambda^*(t,1)  &  \dots  & 0  & 0   \\
\vdots & \vdots &  \vdots &  \ddots  & \vdots & \vdots  \\
0 & 0 & 0 &  \dots &  \lambda^*(t,m+1) & 0  \\
0 & 0 & 0 &  \dots & - \lambda^*(t,m+2) & \lambda^*(t,m+2)  \\
0 & 0 &  0 & \dots &  0 & 0  \\
\end{matrix}
\right]
\eeq
where $\lambda^* (t,l) = (m+3-l) \widehat \gamma (t,l)$.
\hfill \break
(ii) In particular, if $\widehat \gamma (t,l) = \widehat \gamma (l)$ for some strictly positive function
$\widehat \gamma $, then the default counting process $N^*$ is a time-homogeneous Markov chain with state space $E^* = \{0,1,\dots , m+3 \}$ and the following
generator
\beq
\AMC_{N^*} = \left[
\begin{matrix}
- (m+3) \widehat \gamma (0)  & (m+3) \widehat \gamma (0) &  0  &  \dots   & 0  & 0  \\
0 & - (m+2) \widehat \gamma (1) & (m+2) \widehat \gamma (1) &  \dots  & 0  & 0   \\
\vdots & \vdots &  \vdots &  \ddots  & \vdots & \vdots  \\
0 & 0 & 0 &  \dots  & 2 \widehat \gamma (m+1) & 0  \\
0 & 0 & 0 &   \dots  &- \widehat \gamma (m+2) & \widehat \gamma (m+2)  \\
0 & 0 &  0 &  \dots  & 0 & 0  \\
\end{matrix}
\right].
\eeq
\el

\proof We will first show that assumption (\ref{gtg}), combined with Definition \ref{gtgy}, uniquely
specifies the intensities $\lambda ((j_0, j_1,j_2,0),(j_0+1, j_1,j_2,0))$ and $\lambda ((j_0, j_1,j_2,0),(j_0+1, j_1,j_2,1))$.
To this end, we note that equality (\ref{gtg}) and Definition \ref{gtgy} yield
\bde
\lambda ((j_0, j_1,j_2,1),(j_0+1, j_1,j_2,1))  =(m-j_0) \widehat \gamma (t,l) = \gamma (t,l)
\ede
where $l = j_0+j_1+j_2+1$. Using Definition \ref{gtgy}, we get
\bde
\lambda ((j_0+1, j_1,j_2,0),(j_0+2,j_1,j_2,0)) = a_{j_0+1} \gamma (t,l) = \frac{m-j_0-1}{m-j_0}\, (m-j_0) \widehat \gamma (t,l)
= (m-j_0-1) \widehat \gamma (t,l).
\ede
Similarly, by applying once again Definition \ref{gtgy}, we also obtain
\bde
\lambda ((j_0+1, j_1,j_2,0),(j_0+1, j_1,j_2,1)) = b_{j_0+1} \gamma (t,l)= \frac{1}{m+1-j_0} \, \gamma (t,l) = \widehat \gamma (t,l).
\ede
It is now clear that under the assumptions of part (i), the choice of a non-negative function $\widehat \gamma$ fully specifies
the generator of a time-inhomogeneous Markov contagion model $\MM$ of Definition \ref{gtgy}.

 We also note that in that case the
intensity of default depends only on the total number of defaults up to time $t$ and the size of sub-portfolio. Hence
it is rather obvious that the process $N^*$ has the Markov property and its generator $\AMC_{N^*}$ is represented by the matrix
given in the statement of the lemma. Part (ii) is an immediate consequence of part (i).
\endproof

It is obvious that we can easily compute the default counting process $N^*$ from the Markov chain $\MM$. The converse does not
hold, however, that is, the process $\MM$ cannot be recovered from $N^*$, in general.
This is intuitively clear, since $N^*$ does not convey any information about the identities
of defaulted names. By contrast, the Markov chain $\MM$ specifies not only the total number of defaulted
names, but also the default status of the investor, the counterparty, and the reference entity.
The time-homogeneous Markovian contagion model introduced in Lemma \ref{lmm} (ii)
allows for the examination of not only systemic and at first default
contagions, but also of wrong or right way risks. As a matter of fact, in the event of the default
of an obligor, the systemic risk increases (systemic contagion), which automatically impacts the creditworthiness of the non-defaulting obligor
(at-first-default contagion) as well as the
creditworthiness of the reference entity of the credit default swap. Hence the value of the credit
 default swap jumps in the event of a party's default (wrong or right way risks). It is important to notice
that the at-first-default contagion effect cannot be accounted for under
convention $a)$, but only under other conventions introduced in Section \ref{sect2}.
Also, given that all of the companies bear systemic risk, it makes practical sense to assume that the
intensities of default of the non-defaulting companies increase in the event of a default. Consequently, $\widehat \gamma (l)$
is chosen to be an increasing function of $l$. This choice corresponds to the case of a wrong way risk with a negative at-first-default contagion effect
 for the investor, which is consistent with our illustrative case study
of the CDS between Goldman Sachs, the buyer of protection on Lehman Brothers, and AIG, the seller of protection.

\ssc{CVA of a CDS in a Markovian Contagion Model} \label{sect3.3}

To illustrate our methodology, we will henceforth examine the CVA computation for the
credit default swap with bilateral counterparty risk.

\sssc{Risk-Free Credit Default Swap} \label{sect3.1}

Let $(\Om,\G,\Q) $ be a probability space endowed with a filtration $ \FG= (\G_t)_{0 \le t \le T} $,
where $\Q$ represents the risk-neutral probability measure and $T$ stands for
the horizon date of trading activities. We denote by $r$ the {\it short-term rate process} and by $B$
the {\it savings account} process. The next lemma is an immediate consequence of the definition
of a stylized credit default swap (CDS). Recall that $\tau = \tau^3$ denotes the default time of
a reference credit name for which the protection is sought.

\bl
The {\it discounted cash flows} $\Pi(t,T) $ as seen by the buyer at time $t$ of a risk-free CDS
of maturity $T$ and notional amount 1 are
\bde
\Pi(t,T) =\I_{\{ t \le \tau \le T \}} \frac{B_t}{B_{\tau}} (1-R_{\tau} ) -\kappa \int_{t}^{\tau \wedge T} \frac{B_t}{B_{u}} \, du
\ede
where $\tau $ is the default time of the reference entity, $R_{\tau}$ is the {\it recovery rate}, that is, the rate of recovery at time $\tau$ on the assets
of the reference entity after a credit event and $\kappa $ is the fixed CDS spread.
\el

For simplicity, we assume henceforth that the recovery rate $R$ is constant and the interest rate $r$  is null.
By taking the conditional  expectations under $\Q$, we obtain the following
representation for the price of the risk-free CDS, on the event $\{\tau >t\}$,
\be \lab{RP1}
P_t = (1-R) \, \Q[\tau \le T \mid \G_t] -\kappa \, \EE[ \tau \wedge T - t \mid \G_t].
\ee
From the Markov property of the process $\MM$ and formula (\ref{RP1}), it is easy to deduce
that the {\it pre-default price} of the risk-free credit default swap
(that is, the price on the event $\{ \tau^1 \wedge \tau^2 \wedge \tau  > t\}$) can be represented as a function
of $N_t$.  Specifically, equation (\ref{RP1}) becomes, on the event $\{ \tau^1 \wedge \tau^2 \wedge \tau  > t\}$,
\beq
P_t = P_t(N_t) =  (1-R) \, \Q[\tau \le T \mid N_t] -\kappa \, \EE[ \tau \wedge T - t  \mid N_t] .
\eeq

\sssc{CVA of a Credit Default Swap: Analytic Formula} \label{sect3.3}

We will now derive a convenient expression for the CVA of a CDS in a general Markovian contagion model.
For the ease of explanation and conciseness, we assume henceforth that the settlement value, as seen by the
investor in the event of his default, is null, that is, $S^{i,1}_t =0$ (this assumption can be easily relaxed).
Note that formula established in Proposition \ref{SF11} can be combined with any rank $n$ pricing rule introduced in Section \ref{rankn}.
Given the value $N^*_t$ of the default counting process,
the price at time $t$ of the counterparty equivalent contract is denoted as either $P_t^{*,i,2}(N^*_t)$ or $P_t^{*,i,2|N^*_t}$.
We observe that $M^{i,2}_t = P_t^{*,i,2}(N^*_t)$, that is, the MtM value and the replacement cost coincide; this property is clear since both quantities
refer to the same convention $i)$. Note also that the equality $N^*_t=N_t$ holds on the event $\{ \tau^1 \wedge \tau^2 \wedge \tau  > t\}$.

\bp \lab{SF11}
The CVA of a CDS under the settlement convention $i)$ equals, for every $t < T$ on the event $\{ \tau^1 \wedge \tau^2 \wedge \tau  > t\}$,
\beqa \lab{A1}
\CVA^i_t &=& \int_t^T \bigg( \sum_{v = 0}^{m-N_t} P_u(N_t+v+1)\, \Q_{B^1_u}[N_u = N_t+v \mid N_t  \big]
\, d\Q\big[ \tau^1 \leq u,\, \tau^2 \wedge \tau  > \tau^1  \mid N_t \big] \nonumber
\\ \mbox{} && +  \sum_{v = 0}^{m-N_t} P_u(N_t+v+1) \, \Q_{B^2_u}[N_u = N_t+v \mid N_t  \big]
\, d\Q\big[ \tau^2 \leq u,\, \tau^1 \wedge \tau > \tau^2 \mid N_t \big]
\\ \mbox{} && -  \sum_{v = 0}^{m-N_t} \Big[ S^2_{\tau^2}\big( P^{*,i,2}_u (N_t + v+1) \big)  \Big]
\Q_{B^2_u} [N_u = N_t+v \mid N_t  \big] \, d\Q\big[ \tau^2 \leq u,\, \tau^1 \wedge \tau  > \tau^2 \mid N_t \big] \bigg) \nonumber
\eeqa
where $B^1_u= \{  \tau^1 = u,\, \tau^2 \wedge \tau > \tau^1 \}$ and $B^2_u= \{  \tau^2 = u,\, \tau^1 \wedge \tau > \tau^2 \}$
are two families of events from $\G_T$ indexed by the date $u \in [t,T]$.
\ep

\proof
Let us first clarify the meaning of notation used in the statement and the proof of the proposition. For any event $B \in \G_T$
and any $\Q$-integrable random variable $X$, we define the conditional expectation ${\mathbb E}_{\Q_B}[X\mid N_t]$,
where the probability measure $\Q_B$ is defined as follows
\be  \label{bayesxyz}
 \Q_B (D) = {{\Q ( D \cap B)}\over{\Q (B)}}, \quad \forall \, D \in \G_T ,
\ee
that is, $\Q_B$ is the conditional probability measure on $(\Omega , \G_T)$ given the event $B$. It is known that
the following version of the Bayes formula holds (see Lemma A.3 in Collin-Dufresne et al. \cite{CGH})
\be \label{bayesxy}
\EE [ \I_B X \mid N_t ] = \Q [ B \mid N_t] \, {\mathbb E}_{\Q_B} [X \mid N_t].
\ee
Let us now fix some $t<T$. We start the derivation of formula (\ref{A1}) by observing that, under convention $i)$
and the assumption that the settlement value in the event of the investor's default is null,
the cash flows of a general bilateral contract are
\beq
\Pi^i(t,T) &=& \I_{\{ \tau^1 \wedge \tau^2 > T \}} \, \Pi(t,T)
+ \I_{\{ \tau^1 \le T,\, \tau^2 > \tau^1 \}} \Pi(t,\tau^1)
+ \I_{\{ \tau^2 \le T,\, \tau^1 > \tau^2 \}} \, \Big[ \Pi(t,\tau^2) + S^2_{\tau^2} \Big]
\\ &=& \Pi(t,T) -  \I_{\{ \tau^1 \le T,\, \tau^2 > \tau^1 \}}
\Pi(\tau^1, T) -  \I_{\{ \tau^2 \le T,\, \tau^1 > \tau^2 \}} \,  \Pi(\tau^2, T)
+ \I_{\{ \tau^2 \le T,\, \tau^1 > \tau^2 \}} \,  S^2_{\tau^2},
\eeq
where the settlement value $S^2_{\tau^2} = S^2_{\tau^2} \big( P^{*,i,2}_{\tau^2} (N^*_{\tau^2}) \big)$ is given by formula (\ref{sv01}).
Observe that
\beq
\Pi(\tau^2, T)  = \I_{\{ \tau^2 < \tau \}} \Pi(\tau^2, T) + \I_{\{ \tau^2 \ge \tau \}} \Pi(\tau^2, T) =  \I_{\{ \tau^2 < \tau\}} \Pi(\tau^2, T),
\eeq
since the second term is null; similarly, $\Pi(\tau^1, T)  =  \I_{\{ \tau^1 < \tau \}} \Pi(\tau^1, T)$.
We thus obtain
\beq
\Pi^i(t,T) &=&
\Pi(t,T) -  \I_{\{ \tau^1 \le T,\, \tau^2 \wedge \tau   > \tau^1 \}}
 \Pi(\tau^1, T) -  \I_{\{ \tau^2 \le T,\, \tau^1 \wedge \tau  > \tau^2 \}} \,  \Pi(\tau^2, T)
 \\  &&\mbox{} +   \I_{\{ \tau^2 \le T,\, \tau^1 \wedge \tau  > \tau^2 \}} \, S^2_{\tau^2}.
 \eeq
By taking the conditional expectation under the risk-neutral probability measure $\Q$, we obtain,
on the event $\{ \tau^1 \wedge \tau^2 \wedge \tau  > t\}$,
\bde
\widehat {P}^i_t = \EE[ \Pi^i(t,T) \mid N_t ] = \widehat {P}^i_t (N_t)
\ede
or, more explicitly,
\beq
\widehat {P}^i_t (N_t) &=& \EE[ \Pi(t, T) \mid N_t ] -  \EE[\I_{\{ t< \tau^1 \le T,\, \tau^2 \wedge \tau  > \tau^1 \}}  \Pi(\tau^1, T) \mid N_t ]
 \\  &&\mbox{} - \EE[ \I_{\{ t< \tau^2 \le T,\, \tau^1 \wedge \tau  > \tau^2 \}} \,  \Pi(\tau^2, T) \mid N_t ] + \EE[ \I_{\{ t< \tau^2 \le T,\, \tau^1 \wedge \tau  > \tau^2 \}} \, S^2_{\tau^2}  \mid N_t ].
 \eeq
Using the law of total probability and formula (\ref{bayesxy}), we can now represent $\widehat {P}^i_t (N_t)$ as follows
 \beq
\widehat {P}^i_t (N_t)&=&
P_t (N_t) -  \int_t^T {\mathbb E}_{\Q_{B^1_u}} \big[\Pi(u, T) \mid N_t \big] \, d\Q\big[ \tau^1 \leq u,\, \tau^2 \wedge \tau > \tau^1 \mid N_t \big]
 \\ && \mbox{} - \int_t^T {\mathbb E}_{\Q_{B^2_u}} \big[\Pi(u, T) \mid N_t \big]  \, d\Q\big[ \tau^2 \leq u,\, \tau^1 \wedge \tau > \tau^2 \mid N_t \big]
\\ && \mbox{} + \int_t^T {\mathbb E}_{\Q_{B^2_u}}[ S^2_u \big( P^{*,i,2}_u (N^*_u) \big)  \mid N_t \big] \, d\Q\big[ \tau^2 \leq u,\, \tau^1 \wedge \tau > \tau^2 \mid N_t \big].
 \eeq
 To establish the desired formula, it remains to observe that the conditional expectations
\bde
 {\mathbb E}_{\Q_{B^1_u}} \big[\Pi(u, T) \mid N_t \big], \quad  {\mathbb E}_{\Q_{B^2_u}} \big[\Pi(u, T) \mid N_t \big],
 \quad {\mathbb E}_{\Q_{B^2_u}} \big[S^2_u \big( P^{*,i,2}_u (N^*_u) \big) \mid N_t \big]
\ede
only depend upon the number $N_t$ of defaults and they can be represented more explicitly.
For instance, for every $u \in [t,T]$, the first conditional expectation
can be computed by conditioning on the number of defaults in the systemic portfolio (excluding, of course, the reference credit name).
To be more specific, using (\ref{bayesxy}) and the equality
\beq
\EE \big[\Pi(u, T) \mid  N_u = N_t+v , B^1_u , N_t\big] &=& \EE \big[\Pi(u, T) \mid  N^*_u = N_t+v+1 , \tau^1 \wedge \tau > u \big]
\\& =& P_u (N^*_u)= P_u(N_t+v+1) ,
\eeq
we obtain
\beq
{\mathbb E}_{\Q_{B^1_u}} \big[\Pi(u, T) \mid N_t \big] &=& \frac{ \EE [ \I_{B^1_u} \Pi(u, T) \mid N_t ]}{\Q [ B^1_u \mid N_t]}
= \frac{1}{\Q [ B^1_u \mid N_t]} \, \sum_{v = 0}^{m-N_t} \EE [\I_{\{  N_u = N_t+v} \I_{B^i_u} \Pi(u, T) \mid N_t ]
\\ &=& \frac{1}{\Q [ B^1_u \mid N_t]} \, \sum_{v = 0}^{m-N_t} \EE [ \Pi(u, T) \mid N_u = N_t+v, B^1_u, N_t ] \, \Q [ N_u = N_t+v, B^1_u \mid N_t ]
\\ &=& \sum_{v = 0}^{m-N_t} P_u(N_t+v+1) \, \Q_{B^1_u} \big[ N_u = N_t+v \mid N_t  \big],
\eeq
as well as analogous representations for the other two conditional expectations.
To complete the proof, it thus suffices to use the equality $\CVA ^i_t = P_t - \bhP^i_t$ (see Definition \ref{cvai}).
\endproof

It is worth noting that for every $u \in [t,T]$ the quantity $P_u (N^*_t)$ is a deterministic function of $N^*_t$,
and thus the random variable $P_u(N_t+v+1)$ is measurable with respect to the $\sigma$-field $\sigma (N_t)$.

\sssc{CVA of a Credit Default Swap: Discrete-Time Approximation} \label{sect3.3}

Our next goal is to establish a convenient approximation for the integral in the right-hand side of formula (\ref{A1}).
To this end, we introduce a tenor of dates $t_0 = 0 < t_1 < \cdots < t_M = T$ and we formally postpone a default
from any date $u \in ]t_{j}, t_{j+1} ]$ to the date $t_j$. This leads to the following useful approximation result for the CVA.

\bp \lab{SF1}
The CVA of a CDS under the settlement convention $i)$ can be approximated in a tenor of dates $t=t_0, t_1, \ldots, t_M$ as follows
\beq
\lefteqn{ \CVA^i_t \simeq
\sum_{j=0}^{M-1} \bigg(
\sum_{v = 0}^{m-N_t} P_{t_{j+1}}(N_t+v+1) \, \Q_{\hat B^1_{t_{j+1}}} \big[ N_{t_{j+1}} = N_t+v \mid N_t  \big]\,
\Q\big[ \tau^1 \in ]t_j,{t_{j+1}}],\, \tau^2 \wedge \tau   > \tau^1  \mid N_t \big]}
\\ \mbox{} && +  \sum_{v = 0}^{m-N_t} P_{t_{j+1}}(N_t+v+1) \, \Q_{\hat B^2_{t_{j+1}}}  \big[N_{t_{j+1}} = N_t+v \mid N_t \big]\,
\Q \big[ \tau^2  \in ]t_j,{t_{j+1}}],\, \tau^1 \wedge \tau > \tau^2  \mid N_t \big]
\\ \mbox{} && -  \sum_{v = 0}^{m-N_t} S^2_{t_{j+1}}\Big( P^{*,i,2}_{t_{j+1}} \big(N_t + v + 1\big) \Big)
\Q_{\hat B^2_{t_{j+1}}}  \big[N_{t_{j+1}} = N_t+v \mid N_t \big] \, \Q \big[ \tau^2  \in ]t_j,{t_{j+1}}],\, \tau^1 \wedge \tau   > \tau^2  \mid N_t \big] \bigg)
\eeq
where we denote $\hat B^1_{t_{j+1}}= \big\{ \tau^1 \in ]t_j,{t_{j+1}}],\, \tau^2 \wedge \tau   > \tau^1 \big\}$ and
$\hat B^2_{t_{j+1}}= \big\{  \tau^2  \in ]t_j,{t_{j+1}}],\, \tau^1 \wedge \tau   > \tau^2 \big\}.$
\ep

\proof  Formula (\ref{A1}) yields
\beq
\lefteqn{\CVA^i_t =
\sum_{j=0}^{M-1} \int_{t_j}^{t_{j+1}}\bigg( \sum_{v = 0}^{m-N_t} P_u(N_t+v+1) \, \Q_{B^1_u} \big[N_u = N_t+v \mid N_t  \big]
\, d\Q\big[ \tau^1 \leq  u,\, \tau^2 \wedge \tau  > \tau^1  \mid N_t \big]} \nonumber
\\ \mbox{} && +  \sum_{v = 0}^{m-N_t} P_u(N_t+v+1) \, \Q_{B^2_u}[N_u = N_t+v \mid N_t  \big]
\, d\Q\big[ \tau^2 \leq u,\, \tau^1 \wedge \tau > \tau^2 \mid N_t \big]
\\ \mbox{} && -  \sum_{v = 0}^{m-N_t} \Big[ S^2_{\tau^2}\Big( P^{*,i,2}_u (N_t + v+1) \Big)  \Big]
\Q_{B^2_u} [N_u = N_t+v \mid N_t  \big] \, d\Q \big[ \tau^2 \leq u,\, \tau^1 \wedge \tau  > \tau^2 \mid N_t \big] \bigg) . \nonumber
\eeq
Therefore, by assuming that defaults may only be observed at tenor dates $t=t_1, \ldots, t_M$
(that is, by formally postponing the default event from any time $u \in ]t_{j}, t_{j+1} ]$ to the date $t_{j+1}$), we obtain
 the desired formula.
\endproof

\ssc{CVA Algorithms for a CDS in a Markovian Contagion Model} \label{sect3.4}

Our final goal is to develop a fairly general
algorithm for the CVA of a credit default swap under alternative settlement conventions and margin agreements.
Let us first emphasize that we consider the credit default swap under the standard assumption that the contract
value is null at its inception. In other words, we assume that the obligors agree on a CDS spread value which
makes the contract valueless at time 0. Under this postulate, the
CVA, which is equal to the difference between the price of the risk-free contract and the price of the contract
with bilateral counterparty risk, is simply equal to the price of the corresponding risk-free CDS.

\sssc{Algorithm Without Systemic Risk}

Let us first examine a suitable numerical procedure in the absence of systemic risk.
We start be noting that, with the exception of convention $a)$, the computation of the CVA introduces a convolution.
This convolution is manifested in the formulae for the CVA established in Propositions \ref{SF11} and \ref{SF1} through the presence of
the settlement value in the event of the counterparty's default, that is, $ S^2( P^{*,i,2})$. In effect,
no matter which specification in terms of margin agreements and segregation is chosen, that is,
for either of formulae  (\ref{5A}), (\ref{7A}), (\ref{8A}), (\ref{9A}), or (\ref{sv01}), the settlement
value $S^2( P^{*,i,2})$ is a function of the price $P^{*,i,2}$ of the equivalent contract in the event of the counterparty's default,
where $P^{*,i,2}$ is in turn a function of the price $P^{*,i,3}$ of the second counterparty equivalent contract in the event of default of the
first equivalent counterparty, and so on.

In the present set-up, the computational challenge due to this convolution can be tackled by means of a recursive procedure performed on a suitably defined tree.
Within this tree, the node $(k +1,k + 1)$ is used for storing the price of the $k$th equivalent contract at time $t_k$. For instance, the price of
the first equivalent contract at time $t_1$, in the event of default of the counterparty between $]t_0, t_1]$ (or `at time $t_1$', for short), is stored
at the node $(2,2)$, whilst the price of the $n$th equivalent contract at time $T=t_M$, in the event of default
of the $(n-1)$th equivalent contract `at time $t_{M}$', is stored at the node $(n+1, M+1)$.

At each step of the recursive  procedure, the tree is constructed by making use of the values of the $k$th equivalent contract price in the event of default
of the $(k-1)$th equivalent contract at time $t_k, t_{k+1}, \ldots, t_{M}$, which are stored at nodes
$( k +1, k + 1), ( k+1, k+2),  \ldots, ( k+1, M + 1)$,
so as to compute the price of the $(k-1)$th equivalent contract in the event of default of the $(k-2)$th equivalent contract at time $t_{k-1}$. The latter value
is then stored at the node $( k , k )$. The values, which are calculated through a similar step, at the nodes
$(k , k), ( k  , k + 1),  \ldots, ( k  , M + 1),$
are subsequently used for the computation of the $(k-2)$th equivalent contract price in the event of default of the $(k-3)$th counterparty at time $t_{k-2}$.
The latter value is then stored in the node $ ( k  - 1 , k - 1)$, and so on until the price of the bilateral credit default swap at time 0, at node $(1,1)$.

Clearly, this computation must be initiated by first calculating the values of the price of the $n$th counterparty equivalent contract at nodes
$( n + 1, n+1), ( n + 1, n+2), \ldots , ( n + 1, M + 1)$,
which not only depend on the selected rank $n$ pricing rule, but also on the spread value.
Consequently, under the assumption that the CDS is worthless at its inception, the previously described iteration is run until a spread value
for which the price of the credit default swap is sufficiently close to zero is found. This can be easily achieved,
by adjusting the spread value all along the numerical procedure.
As soon as a satisfactory spread value is found then,  in order to obtain the CVA, it suffices to compute the value of the risk-free contract at inception
for the final value of the spread.
Note that the recursive steps require the knowledge of conditional probabilities of default, which, in practice, are computed  ahead of the tree construction.
This is an important feature of the pricing engine because it shows that it is readily applicable to any models of correlated default times, such as:
copula-based models or affine models.

\sssc{Algorithm with Systemic Risk}

For computation of the CVA in the presence of systemic risk, the algorithm is essentially the same as the one just previously described.
However,  we now need to account for the dependency of the
settlement values on the level of systemic risk, that is, in our setting, on the number of systemic defaults.
A simple and natural way of extending the previous procedure relies on adding to the tree a third dimension, characterizing
the number of systemic defaults. In addition, a significant improvement of computational efficiency can be achieved by
restricting the summation with respect to $v$ in approximate formula of Proposition \ref{SF1} to a reduced number of possible systemic defaults.
To this end, for every $j=0,1, \dots, M-1$ and $x >j$, we denote by $N_{j,x+1}^{l}$ a
set of numbers of defaults such that the random variable $N^*_{t_{x+1}}$ takes values in this set with a high probability, given that $N^*_{t_j} =l$.
Then the iterations are run by calculating the relevant quantities  on the tree, say at node $(k,j,l)$,
as a function of the quantities computed at nodes $(k+1, x, v )$ for every $x>j$ and any number $v \in N_{j,x+1}^{l}$ of systemic defaults.
We now present the procedure more formally in the case of the time-homogeneous Markovian contagion model of Lemma  \ref{lmm}(ii).
We recall that we refer to the creditworthiness  of a company as a function of its default intensity, which in turn is a function
of the number of systemic defaults. Hence we will write $c^1_{t_j} = \widehat \gamma (N^*_{t_j})$ and $c^2_{t_j} = \widehat \gamma (N^*_{t_j})$ for the
creditworthiness of the investor and the counterparty at time $t_j$, respectively.
In addition, we will denote by $ \tau^{2, k}$ the moment of default of the $k$th counterparty equivalent contract, and
 by $c^{2,k}_{t_j} $ the creditworthiness of the $k$th counterparty, at time $t_j$.
Also, the price at time $t_j$ of the $k$th counterparty equivalent contract, given the number $l$ of systemic defaults, will be denoted as $P_{t_j}^{*,i,k|l}$.

It is important to note that the creditworthiness of the $k$th counterparty equivalent contract depends not only on the choice of a particular convention $i)$, but also on whether or not the equivalent counterparties trigger systemic effects. Under the Markov chain specification of Section \ref{sect3.3}, it is implicitly assumed that it is not the case (see also Remark \ref{rem-bb} for a more sophisticated modeling approach). The random time $\tau^{2, k}$  can be obtained by running an independent copy of model of Definition
\ref{gtgy}, started at time $\tau^{2,k-1}$ of default of the $(k-1)$th counterparty and with the initial number of defaults in the systemic portfolio equal to the observed number of defaults in this portfolio at time $\tau^{2,k-1}$. To be more specific,
let us represent the joint dynamics of defaults for the systemic portfolio, the investor, and the $k$th counterparty by a continuous-time Markov chain $\MM = (\MM_t)_{t \in [\tau^{2,k-1},T]}$
with the state space
\beq
 E = \{ (j_0,0,0,0),(j_0,1,0,0),(j_0,0,1,0),(j_0,0,0,1), (j_0,1,1,0), \\ (j_0,1,0,1),(j_0,0,1,1),(j_0,1,1,1),\, j_0 =0,1,\dots , m \},
\eeq
where the generic state $(j_0, j_1,j_2, j_3)$ has the following interpretation: \hfill \break
(a) $j_0$ is the number of defaults in systemic portfolio ($j_0 \ge N_{\tau^{2,k-1}}$), \hfill \break
(b) $j_1$ is the default indicator for the investor, \hfill \break
(c) $j_2$ is the default indicator for the $k$th counterparty, \hfill \break
(d) $j_3$ is the default indicator for the reference entity.

For instance, under convention $b)$ we will have that
\beq
&&\lambda ((j_0,0,0,0),(j_0+1,0,0,0)) =  a_{j_0} \gamma (t,j_0+1), \\
&&\lambda ((j_0,0,0,0),(j_0,1,0,0)) = \hat \gamma_1 (t,j_0+1), \\
&&\lambda ((j_0,0,0,0),(j_0,0,0,1)) = b_{j_0} \gamma (t,j_0+1 ), \\
&&\lambda ((j_0,0,0,0),(j_0,0,1,0)) =\hat \gamma_2( l_0 ),
\eeq
where $ \hat \gamma_1 (t, j_0+1) =\hat \gamma_2 (t, j_0+1)  = \hat \gamma (j_0+1)
= b_{j_0} \gamma (t, j_0+1 ) $ under the specification of Lemma \ref{lmm}(ii) and
$l_0$ stands for the initial number of systemic defaults, that is, $N^*_0 = l_0$.

We are in a position to describe in some detail our pricing algorithm.
We assume convention $i)$ for some $i = b, c, c'$ and the investor's rank $n$ pricing rule.
Recall that the approximation for the CVA of a credit default swap in a tenor of dates $t=t_0, t_1, \ldots, t_M$
was established in Proposition \ref{SF1}. We now argue that the following algorithm yields a numerical implementation
of the CVA formula of Proposition \ref{SF1}.

\noindent{\bf Step 1.}  Input: initial value for the spread $\hat{\kappa}^i_0$. Run the following loop until a spread value $\hat{\kappa}^i_0$
such that the contract price is sufficiently small (i.e., as close to zero as desired) is found:
\begin{itemize}
\item{\bf Step 1.1.} Compute the price of the $n$th equivalent contract for every $t_j,\, j = n, \ldots, M-1$
and for each number $l = 0, \ldots, m+2$ of systemic defaults observed by time $t_j$, that is, at nodes
\bde
( n + 1, n+1,l+1), ( n + 1, n+2,l+1), \ldots , ( n + 1, M + 1,l+1).
\ede
For this purpose:
\begin{itemize}
\item under the investor's risk-free rank $n$ pricing rule, use the approximation
\be \lab{GGG}
 P^{*,i,n|l}_{t_j} \simeq (1-R) \, \Q \big[ \tau \le T  \mid N^*_{t_j}=l, \tau > t_j \big]
  - \hat{\kappa}^i_0 \, \EE \big[ \tau \wedge   T - t_j  \mid N^*_{t_j}= l, \tau > t_j  \big],
\ee
\item under the zero rank $n$ pricing rule, set $ P^{*,i,n|l}_{t_j} = 0$.
\end{itemize}
\item{\bf Step 1.2.} Compute the price of the $k$th equivalent contract for every $t_j,\, j = k, \ldots, M-1$
and for each number $l = 0, \ldots, m+2$ of systemic defaults observed by time $t_j$, that is, at nodes
\bde
(k+ 1, k+1,l+1), ( k + 1, k+2,l+1), \ldots , ( k + 1, M + 1,l+1).
\ede
To this end, use the following equation at each
iteration with respect to $k$, where $1 \le k \le j$
\beqa \nonumber
P^{*,i,k |l}_{t_j} &\simeq &  (1-R) \, \Q \Big[ \tau \le T,\, \tau^1 \wedge \tau^{2,k+1} > \tau  \Mid A^{i,k|l}_{j} \Big]
- \hat{\kappa}^i_0 \, \EE \Big[ \tau \wedge \tau^1 \wedge \tau^{2,k+1} \wedge T - t_j  \Mid A^{i,k|l}_{j} \Big]
\\ &&\mbox{} +  \sum_{x=j}^{M-1}  \sum_{v \in N_{j,x+1}^{l}} S^2_{t_{x+1}} \big(  P^{*,i,k+1 |v}_{t_{x+1}} \big)\,
\Q_{\hat B^{2,k+1}_{t_{x+1}}} \Big[  N^*_{t_{x+1}} = v \mid A^{i,k|l}_{j} \Big] \lab{GGG1}
\\ &&\mbox{}\times \Q\Big[t_x < \tau^{2,k+1} \le t_{x+1},\,  \tau \wedge \tau^1 > \tau^{2,k+1} \Mid A^{i,k|l}_{j} \Big] \nonumber
\eeqa
where we denote
$\hat B^{2,k+1}_{t_{x+1}}= \big\{ t_x < \tau^{2,k+1} \le t_{x+1},\,  \tau \wedge \tau^1 > \tau^{2,k+1}\big\}$
and
\beq
&&A^{a',k|l}_{j} = \big\{ t_{j-1}< \tau^{2,k} \leq t_j ,\, \tau \wedge \tau^1 > \tau^{2,k} ,\,
  N^*_{t_j} = l,\, c^1_{t_j} = \widehat \gamma (l),\,  c^{2,k}_{t_j} =  \infty\big\},\\
&&A^{b,k|l}_{j} = \big\{  t_{j-1}< \tau^{2,k} \leq t_j ,\, \tau \wedge \tau^1 > \tau^{2,k} ,\,
   N^*_{t_j} = l,\, c^1_{t_j} = \widehat \gamma (l),\,  c^{2,k}_{t_j} =  \widehat \gamma (l_0)\big\},\\
&&A^{c,k|l}_{j} = \big\{  t_{j-1}< \tau^{2,k} \leq t_j ,\, \tau \wedge \tau^1 > \tau^{2,k} ,\,
   N^*_{t_j} = l,\, c^1_{t_j} = \widehat \gamma (l),\,  c^{2,k}_{t_j} =  \widehat \gamma (l-1)\big\},\\
&&A^{c',k|l}_{j} = \big\{  t_{j-1}< \tau^{2,k} \leq t_j ,\, \tau \wedge \tau^1 > \tau^{2,k} ,\,
   N^*_{t_j} = l,\, c^1_{t_j} = \widehat \gamma (l - 1) =   c^{2,k}_{t_j} \big\},
\eeq
and where $N_{j,x+1}^{l}$ denotes a set of number of defaults such that $N^*_{t_{x+1}}$ takes values
in this set with a high probability given that $N^*_{t_j} = l$. Recall also that $l_0=N^*_0$.
\item{\bf Step 1.3.} Compute the price of the contract at time 0 from the formula
\beqa
\hat P^i_0 &\simeq & (1-R) \, \Q \big[ \tau \le T,\, \tau^1 \wedge \tau^2 > \tau \big]
- \hat{\kappa}^i_0 \ \EE \big[ \tau \wedge \tau^1 \wedge \tau^2 \wedge T   \big]\lab{GGG2}
\\ &&\mbox{} +  \sum_{x=0}^{M-1}  \sum_{v \in N_{0,x+1}^{l_0}} S^2_{t_{x+1}}
\big(  P^{*,i|v}_{t_{x+1}} \big) \, \Q_{\hat B^{2}_{t_{x+1}}} \big[  N^*_{t_{x+1}}
= v \big] \, \Q\big[t_x < \tau^2 \le t_{x+1} ,\, \tau^1 \wedge \tau > \tau^2 \big] \nonumber
\eeqa
where  $\hat B^{2}_{t_{x+1}}= \big\{ t_x < \tau^{2} \le t_{x+1},\,  \tau \wedge \tau^1 > \tau^{2}\big\}$
and $N_{0,x+1}^{l_0}$ denotes a set of number of defaults such that $N^*_{t_{x+1}}$ belongs to this set
with a high enough probability, given that $N^*_{0} = l_0$.
\end{itemize}

\noindent{\bf Step 2.}   Compute the CVA of a CDS at time 0: it is given by $\CVA^i_0 = P_0(\hat{\kappa}^i_0)$.

\vskip 5 pt We will now justify the algorithm by postulating, for concreteness, the convention $b)$.
Note that Step 2 is a straightforward consequence of the definition of the CVA  (recall that $\hat P^b_0\big( \hat \kappa^b_0  \big)=0$)
\bde
\CVA^b_0 = P_0\big( \hat \kappa^b_0  \big) - \hat P^b_0\big( \hat \kappa^b_0  \big) =P_0(\hat{\kappa}^b_0).
\ede
For Step 1, we will only focus on derivation of equation (\ref{GGG1}), since formula (\ref{GGG2})
can be obtained using analogous arguments,  and formula (\ref{GGG}) is clear. Equality (\ref{GGG1})
 hinges on the following two simplifications:
(i) the postulate that defaults can only occur at the tenor dates,
(ii) the summation of conditional expectations of settlement values over a reduced set of possible defaults.

Let us write
\bde
A^{b,k|l}_{j} = \big\{  t_{j-1}< \tau^{2,k} \leq t_j ,\, \tau \wedge \tau^1 > \tau^{2,k} ,\, N^*_{t_j} = l,\,
c^1_{t_j} = \widehat \gamma (l),\, c^{2,k}_{t_j} = c^2_0 = \widehat \gamma (0) \big\}.
\ede
We start by noting that the price $P^{*,b,k |l}_{t_j}$ of the $k$th equivalent investor contract given that $N^*_{t_j}=l$, equals
\beq
P^{*,b,k | l}_{t_j} &=& \EE \bigg[\I_{\{ \tau \le T,\, \tau^1 \wedge \tau^{2,k+1} > \tau \}} (1-R)
 -\int_{t_j}^{T \wedge \tau^1 \wedge \tau \wedge \tau^{2,k+1}} \hat{\kappa}^b \, du
\\ && \mbox{} + \I_{\{ \tau^{2,k+1}  \le T,\,  \tau^1 \wedge \tau > \tau^{2,k+1}\}}  S^2_{\tau^{2,k+1}} \big( P^{*,b,k+1}_{\tau^{2,k+1}} \big) \Mid  A^{b,k|l}_{j} \bigg]
\\ &=&  (1-R) \, \Q \Big[\tau \le T,\, \tau^1 \wedge \tau^{2,k+1} > \tau \Mid  A^{b,k|l}_{j} \Big]
\\ &&\mbox{}- \hat{\kappa}^b_0 \, \EE \bigg[ \int_{t_j}^{T} \I_{\{ u < \tau \wedge \tau^1 \wedge \tau^{2,k+1} \}} \, du \Mid  A^{b,k|l}_{j} \bigg]
\\ && \mbox{} + \EE \Big[ \I_{\{ \tau^1 \wedge \tau > \tau^{2,k+1}\}} \, \I_{\{ \tau^{2,k+1}  \le T\}} \,
S^2_{\tau^{2,k+1}} \big( P^{*,b,k+1}_{\tau^{2,k+1}} \big) \Mid  A^{b,k|l}_{j} \Big].
 \eeq
This means that $P^{*,b,k |l}_{t_j} = (1-R) J  - \hat{\kappa}^b_0 I + K$ where we denote
\bde
I = \EE \Big[ \tau \wedge \tau^1 \wedge \tau^{2,k+1} \wedge T - t_j  \Mid  A^{b,k|l}_{j} \Big]
\ede
and
\bde
J = \Q \Big[ \tau \le T, \, \tau^1 \wedge \tau^{2,k+1} > \tau \Mid  A^{b,k|l}_{j} \Big].
\ede
We thus need to approximate the term denoted by $K$. For this purpose, we observe that
\beq
K &=& \EE \Big[ \I_{\{ \tau \wedge \tau^1 > \tau^{2,k+1}\}} \, \I_{\{ \tau^{2,k+1}  \le T\}} \,
S^2_{\tau^{2,k+1}} \big( P^{*,b,k+1}_{\tau^{2,k+1}} \big) \Mid  A^{b,k|l}_{j} \Big]
\\ \mbox{} &=& \int_{t_j}^T {\mathbb E}_{\Q_{B^{2,k+1}_u}} \Big[   S^2_u \big( P^{*,b,k+1}_u \big)\Mid  A^{b,k|l}_{j} \Big]
\, d\Q\Big[ \tau^{2,k+1} \le u ,\, \tau^1 \wedge \tau > \tau^{2,k+1}\Mid  A^{b,k|l}_{j} \Big]
\\ \mbox{} &=& \sum_{x=j}^{M-1} \int_{t_x}^{t_{x+1}}  {\mathbb E}_{\Q_{B^{2,k+1}_u}} \Big[   S^2_u \big( P^{*,b,k+1}_u \big)\Mid  A^{b,k|l}_{j} \Big]
\, d\Q\Big[ \tau^{2,k+1} \le u ,\, \tau^1 \wedge \tau > \tau^{2,k+1}\Mid  A^{b,k|l}_{j} \Big]
\eeq
where  $B^{2,k+1}_{u} = \{ \tau^{2,k+1} = u , \tau^1 \wedge \tau > \tau^{2,k+1} \}.$

If we assume that defaults can only occur at the tenor dates (that is, equivalently,
if we forcibly postpone default from any time $u \in ]t_{i-1}, t_i ]$ to the date $t_i$),
we obtain the following approximations for $K$
\beq
K &\simeq & \sum_{x=j}^{M-1} \int_{t_x}^{t_{x+1}}  {\mathbb E}_{ \Q_{ \hat B^{2,k+1}_{t_{x+1}} } }  \Big[   S^2_{t_{i+1}} \big(  P^{*,b,k+1}_{t_{i+1}} \big)\Mid  A^{b,k|l}_{j} \Big]
 \, d\Q\Big[ \tau^{2,k+1} \le u ,\, \tau^1 \wedge \tau > \tau^{2,k+1} \Mid  A^{b,k|l}_{j} \Big]
\\ \mbox{} &\simeq & \sum_{x=j}^{M-1} {\mathbb E}_{\Q_{\hat B^{2,k+1}_{t_{x+1}}}} \Big[ S^2_{t_{x+1}} \big( P^{*,b,k+1}_{t_{x+1}} \big)\Mid  A^{b,k|l}_{j}  \Big]
\Q\Big[t_x < \tau^{2,k+1} \le t_{x+1} ,\,  \tau^1 \wedge \tau > \tau^{2,k+1} \Mid  A^{b,k|l}_{j} \Big].
\eeq
Finally, the conditional expectations of settlement value can be approximated as follows
\bde
{\mathbb E}_{\Q_{B^{2,k+1}_{t_{x+1}}}}  \Big[   S^2_{t_{x+1}}\big(  P^{*,b,k+1}_{t_{x+1}} \big) \Mid  A^{b,k|l}_{j}  \Big] \simeq
\sum_{v \in N_{j,x+1}^{l}} S^2_{t_{x+1}} \big(  P^{*,b,k+1 \mid v }_{t_{x+1}} \big) \Q_{\hat B^{2,k+1}_{t_{x+1}}  }  \Big[  N^*_{t_{x+1}} = v \Mid  A^{b,k|l}_{j} \Big]
\ede
where $N_{j,x+1}^{l}$ denotes a set of number of defaults such that $N^*_{t_{x+1}}$
belongs to this set with a high probability (say 95$\%$) given that the number of systemic
defaults at time $t_j$ equals $l$. Modifications of this algorithm from convention $b)$ discussed here
to conventions $c)$ and $c')$ are straightforward.

\brem \lab{rem-bb}
Under the Markov chain specification, as described in Section \ref{sect3.3}, it is implicitly assumed that the defaults of
the successive equivalent contracts do not trigger systemic effects. One can nevertheless account for
such systemic defaults by means of a rather straightforward extension of the model.
Indeed, it suffices to consider the existence of a {\it systemic portfolio of potential counterparties},
which contains the $n-1$ potential counterparties of the contract investor,
and to hypothesize that the equivalent counterparty is always the first-to-default among this basket.
In other words, if, at time $t$, the counterparty has defaulted and $k-1$ subsequent equivalent counterparties
went bankrupt as well, then the $k$th equivalent counterparty is the first-to-default entity
among the remaining $n - 1 - k $ entities within the systemic basket of potential counterparties.
In this setting, the joint dynamics of defaults for the systemic portfolio, the systemic portfolio of potential
counterparties, the investor, the counterparty, and the reference entity can be represented by a continuous-time Markov chain $\MM = (\MM_t)_{t \in [0,T]}$
with the state space $E$ given by
\beq
 E = \big\{ (i,k,0,0,0),(i,k,1,0,0),(i,k,0,1,0),(i,k,0,0,1), (i,k,1,1,0), (i,k,1,0,1), \\ (i,k,0,1,1),(i,k,1,1,1),\, i \in \{0,1,\dots , m\} ,\, k \in \{0,1,\dots , n-1 \} \big\}
\eeq
with the following interpretation of a generic state $(i,k,j_1,j_2,j_3)$: \hfill \break
(a) $i$ is the number of defaults in some systemic portfolio of entities, \hfill \break
(b) $k$ is the number of defaults in a systemic portfolio of potential counterparties, \hfill \break
(c) $j_1$, $j_2$ and $j_3$ are the default indicators for the investor, the counterparty and the reference entity. \hfill \break
This implies, in particular, that $l = i + k+ j_1 + j_2 + j_3$ is the total number of defaults.
 Note also that the sum $i+k$ formally corresponds to the variable $j_0$ in Section \ref{sect3.2}.

Let us now explicitly write the transition intensities in the event that the counterparty has defaulted. We only consider conventions $b)$, $c)$ and $c')$,
since under conventions $a)$ and $a')$ the at-first-default contagion effect does not appear.

\noindent {\bf Convention $b)$}.$\, $  Under convention $b)$,
the defaults within the systemic portfolio of potential counterparties impact the credit rating transitions of each firm in our model,
with the exception of the potential counterparties, of which transition intensities are constant and equal to the credit quality of the counterparty,
at contract inception. We have
\beq
&&\lambda ((i,k,0,1,0), (i+1,k,0,1,0)) =  a_{i} \gamma (t,i+k+1), \\
&&\lambda ((i,k,0,1,0),(i,k,1,1,0)) = \hat \gamma_1 (t,i+k+1), \\
&&\lambda ((i,k,0,1,0),(i,k,0,1,1)) = b_{i} \gamma (t,i+k+1 ), \\
&&\lambda ((i,k,0,1,0),(i,k+1,0,1,0)) =\hat \gamma_2 ( l_0 ),
\eeq
where $l_0\in \{0,1,\dots ,m\}$ is the initial value of $N^*$, that is, $l_0 = N^*_0$.

\noindent {\bf Convention $c)$}.$\, $ Under this convention, the $k$th equivalent counterparty is chosen in such a way
that its credit quality equals the credit quality of the $(k-1)$th equivalent counterparty, just before default.
This implies that the credit quality of the potential counterparties only depend on the number of defaults $i$
within the systemic portfolio. We obtain
\beq
&&\lambda ((i,k,0,1,0), (i+1,k,0,1,0)) =  a_{i} \gamma (t,i+k+1 ), \\
&&\lambda ((i,k,0,1,0),(i,k,1,1,0)) = \hat \gamma_1(t, i+k+1), \\
&&\lambda ((i,k,0,1,0),(i,k,0,1,1)) = b_{i} \gamma (t,i+k+1 ), \\
&&\lambda ((i,k,0,1,0),(i,k+1,0,1,0)) =\hat \gamma_2 (t,i ) .
\eeq
\noindent {\bf Convention $c')$}.$\, $ Finally, under convention $c')$, we take as a reference the price of an equivalent contract between two
firms that have credit qualities the same as those of the obligors of the initial contract just prior to default.
We thus establish the credit rating transition of the investor
by discarding the first, as well as the subsequent, `contagion effects at first default'
\beq
&&\lambda ((i,k,0,1,0), (i+1,k,0,1,0)) =  a_{i} \gamma (t,i+k+1), \\
&&\lambda ((i,k,0,1,0),(i,k,1,1,0)) = \hat \gamma_1 (t,i ), \\
&&\lambda ((i,k,0,1,0),(i,k,0,1,1)) = b_{i} \gamma (t,i+k+1), \\
&&\lambda ((i,k,0,1,0),(i,k+1,0,1,0)) =\hat \gamma_2 (t, i) .
\eeq
\erem

\ssc{Numerical Implementations and Conclusions} \label{sect3.5}

We conclude the paper by presenting a few stylized examples of numerical simulations. Our goal is here to draw some conclusions
regarding the proposed methodology and the impact of alternative settlement conventions on the CVA values, rather than to
calibrate the model to market data.  It should be
mentioned that we work here within a slightly extended framework
in order to examine the effects of different scenarios of credit qualities for the investor, the counterparty,
the reference entity, and the systemic companies. More specifically, we simulated the credit
default swap spreads for several selected values $(M_I, M_{R_2}, M_R)$
of the risk-neutral intensities of default of the investor, the counterparty, and the reference entity,
expressed here in terms of the risk-neutral default intensity of a firm from the cohort of systemic companies.
In our illustrative examples reported below, we consider the set $(M_I, M_{R_2}, M_R)=(1, 0.8, 1.3)$
meaning that, at the contract's inception,
the risk-neutral default intensity of the investor, the counterparty and the reference entity are equal to 1, 0.8 and 1.3
times the risk-neutral default intensity of any firm in a homogeneous portfolio of systemic companies.
This stylized case is thus aimed to describe a situation where
the investor has the same credit quality as the systemic
companies, whilst the credit quality of the counterparty is set at a
better than the average value, as indeed one would typically expect from a protection seller.
By contrast, we postulate that the credit quality reference entity is below the benchmark represented by
any of systemic companies. To be more specific, we work under the assumption that the reference entity has the
risk-neutral default intensity that exceeds by 30$\%$ the systemic average.

In Tables \ref{data1} and \ref{data2}, we report numerical results obtained within this framework for spreads of an
uncollateralized, risk-free rank 3 credit default swap with maturity three years, for two different values of the initial
yearly risk-neutral survival probability of a systemic company.
The recovery rate of the reference entity is set at 0.45 and the recovery rate
of the counterparty is assumed to range from 0.01 to 0.9.
Our procedure uses a tree with $144$ nodes and a standard minimization function.
The total number of systemic companies is chosen to be 10, and we assume that
each systemic default triggers a 150 percent increase of default intensities
of surviving systemic companies.

We first present in Table \ref{data1} simulation results obtained for the risk-neutral yearly probability of survival equal to 0.95.
Under this assumption, the CDS spreads under settlement conventions $a'$), $b$) and $c$) are manifestly very
close to each another, whereas the spreads under convention $a$) (convention $c'$), resp.) appear to be consistently
lower (higher, resp.) than the former. Also, in accordance with the fact that the values of the contracts for the
investor go down concurrently with the decrease of the counterparty's recovery rate $R_2$,
it appears that each spread under convention $i$) for $i=  a, a', b, c, c'$,
becomes lower than the risk-free spread for any value of the recovery rate below 0.4.
One can also notice that the {\it range} between spreads, that is, the relative difference between
the maximum and minimum values across alternative settlement conventions, is rather substantial,
since it starts from around 50 basis points when $R_2=0.1$ and rises to around 500 basis points when
$R_2$ attains the highest considered level of 0.9. The range is given in basis points.

\begin{table}[!h]
\begin{center}
\begin{tabular}{|c|c|c|c|c|c|c|}
\hline
$R_2$&$a$ &$a'$ &$b$&$c$&$c'$&Range\\
\hline
0.9&	0.0558986&	0.0586499&	0.0587415&	0.0588333	&0.0592929&	525\\
0.8& 	0.0553826 &	0.0578922&	0.0579374&	0.0579374& 0.0583900&	461 \\
0.7&	0.0549956&	0.0571009&	0.0571009&	0.0571009& 0.0575024&	383 \\
0.6&	0.0545227&	0.0563969&	0.0563087&	0.0563087&	0.0565727&	344 \\
0.5&	0.0540927&	0.0556140&	0.0555271&	0.0554837&	0.0557005&	281 \\
0.4&	0.0535767&	0.0548324&	0.0547467&	0.0546612&	0.0548320&	234 \\
0.3&	0.0531467&	0.0540601&	0.0539757&	0.0538913&	0.0540597&	172 \\
0.2&	0.0526307&	0.0532475&	0.0532475&	0.0531643&	0.0532473&	117 \\
0.1&	0.0522007&	0.0524454&	0.0524454&	0.0524454&	0.0524454&	47 \\
0.01&	0.0517707&	0.0517707&	0.0517707&	0.0517708&	0.0517708&	0 \\
\hline
\end{tabular}
\end{center}
\caption{CDS spreads under the risk-free rank 3 pricing rule with the initial yearly survival probability 0.95.
The risk-free spread equals 0.0550386.}
\label{data1}
\end{table}

Obviously, the relative values of CDS spreads under each settlement convention depend on
many factors underpinning our general framework and thus the features observed in Table \ref{data1} may fail to persist
when the inputs used for numerical simulations are modified.
For the sake of comparison, we report in Table \ref{data2} results
corresponding to the case of the initial  yearly risk-neutral
survival probability of a systemic firm equal to 0.8.  It is worth noting
that, for a fixed recovery rate $R_2$, the spread under convention $a$) is now higher than the spreads under convention $a'$), $b$) and $c$),
and close to the spread under convention $c'$). The pattern of behavior of the range, when the recovery rate
$R_2$ varies, is also distinctly different for the two cases under study; specifically, it appears to be increasing in $R_2$
in the former case, whereas it is hump-shaped in the latter.

\begin{table}[!h]
\begin{center}
\begin{tabular}{|c|c|c|c|c|c|c|}
\hline
$R_2$&$a$ &$a'$ &$b$&$c$&$c'$&Range\\
\hline
0.9&	0.3783901&	0.3689304&	0.3679216	&0.3672030	&0.3806145	&305 \\
0.8&	0.3691154&	0.3594550&	0.3579807&	0.3561628&	0.3677103&	364 \\
0.7&		0.3595855&	0.3500340&	0.3481881&	0.3456719&	0.3553264	&403 \\
0.6&		0.3497152&	0.3405625&	0.3385670&	0.3355913&	0.3435878	&421 \\
0.5&		0.3395896&	0.3310335&	0.3290292&	0.3260089&	0.3323762	&417 \\
0.4&		0.3291237&	0.3215384&	0.3196544&	0.3167825&	0.3216704	&390\\
0.3&		0.3182323&	0.3120790&	0.3104333&	0.3079474&	0.3114960	&334\\
0.2&		0.3070857&	0.3025274&	0.3012865&	0.2994623&	0.3017434	&255\\
0.1&		0.2954285&	0.2930051&	0.2923183&	0.2912907&	0.2924285	&142\\
0.01&		0.2847073&	0.2844293&	0.2843182&	0.2842071&	0.2843181	&18\\
\hline
\end{tabular}
\end{center}
\caption{CDS spreads under the risk-free rank 3 pricing rule with the initial yearly survival probability 0.8.
The risk-free spread equals 0.4356549.}
\label{data2}
\end{table}

To study the impact of the choice of the pricing rule, we compare in Table \ref{data3} the spreads generated using the risk-free rank $n$ rule,
for the values of $n$ ranging from 1 to 4, and in the situation when the recovery rate of the counterparty equals to 0.4. When the
initial yearly risk-neutral survival probability equals 0.95, the discrepancies tend to be negligible
from rank 2 onwards, that is, we usually do not observe any significant difference between CDS spreads for
ranks 2, 3 and 4. This illustrates that computations for the rank $n=2$ ought to give precise enough results in a typical day-to-day
practice. Intuitively, this is due to the fact that the default event of the consecutive counterparty has small probability
and thus it can be neglected.

\begin{table}[!h]
\begin{center}
\begin{tabular}{|c|c|c|c|c|c|c|c|}
\hline
Rank&$a$ &$a'$ &$b$&$c$&$c'$&Range\\
\hline
1&	0.0535767 &	0.0548324	&0.0535901	&0.0535901	&0.0535901	&234\\
2&	0.0535767	&0.0548324	&0.0547467	&0.0546612	&0.0548747	&234\\
3&	0.0535767	&0.0548324	&0.0547467	&0.0546612	&0.0548320	&234\\
4&	0.0535767	&0.0548324	&0.0547467	&0.0546612	&0.0548320	&234\\
\hline
\end{tabular}
\end{center}
\caption{CDS spreads under the risk-free rank $n$ pricing rule for $n = 1, 2, 3, 4$ with the initial yearly survival probability 0.95 and
the recovery rate $R_2 = 0.4.$}
\label{data3}
\end{table}

When we modify the set-up by bumping up the level of the systemic risk, this approximation may not hold anymore, however.
To illustrate this claim, we display in Table \ref{data4} CDS spreads obtained under the postulate that
the  yearly risk-neutral survival probability equals 0.8. Notice that the spreads under risk-free rank 2
and risk-free rank 3 pricing rules are now significantly different for some settlement conventions,
whilst results for risk-free rank 3 and 4 pricing rules are still close to one another.

\begin{table}[!h]
\begin{center}
\begin{tabular}{|c|c|c|c|c|c|c|}
\hline
Rank&$a$ &$a'$ &$b$&$c$&$c'$&Range\\
\hline
1&	0.3291237&	0.3215384&	0.3291373&	0.3291373	&0.3291373&	236\\
2&		0.3291237&	0.3215384&	0.3202196&	0.3184684&	0.3232579&	335\\
3&		0.3291237	&0.3215384&	0.3196544&	0.3167825&	0.3216704&	390\\
4&		0.3291237	&0.3215384	&0.3196544	&0.3165952	&0.3214802	&396\\
\hline
\end{tabular}
\end{center}
\caption{CDS spreads under the risk-free rank $n$ pricing rule for $n=1,2,3,4$ with the initial yearly survival probability 0.8
and the recovery rate $R_2 = 0.4.$}
\label{data4}
\end{table}

Finally, we compare in Table \ref{data5} the prices of the spreads under the zero rank $n$ and risk-free rank $n$ pricing rules.
For our stylized example, the variations appear to be fairly negligible and, as expected, they tend to go down when
the rank increases.

\begin{table}[!h]
\begin{center}
\begin{tabular}{|l|c|c|c|}
\hline
Rank/convention & $b$ & $c$ & $c'$ \\
\hline
Risk-free rank 2& 0.0547467 &	0.0546612&	0.0548747 \\
Zero rank 2& 0.0546610&	0.0544902&	0.0547031  \\
\hline
Risk-free rank 3& 0.0547467&	0.0546612&	0.0548320\\
Zero rank 3& 0.0547467&	0.0546612&	0.0548747 \\
 \hline
Risk-free rank 4&0.0547467&	0.0546612&	0.0548320\\
Zero rank 4& 0.0547467&	0.0546612&	0.0548320 \\
\hline
\end{tabular}
\end{center}
\caption{CDS spreads under the risk-free and zero rank $n$ pricing rules with the initial yearly survival probability 0.95.}
\label{data5}
\end{table}

\vfill \eject
Results of further simulations, which are not reported here,  show significant variations in the pricing of the credit default swap
for different choices of conventions, systemic risk, and margin agreements, and thus they support the view that
when we account for the bilateral counterparty risk and systemic risk, as well as various settlement conventions,
we obtain the CVA values, which may significantly differ from the one obtained through more
traditional pricing methods. Needless to say that fully practical implementations of alternative conventions and
credit contagion models put forward in this paper are yet to be examined, preferably within a professional environment.

\vskip 15 pt
\noindent {\bf Acknowledgement.}
The research of Marek Rutkowski was supported under Australian Research Council's Discovery Projects funding scheme
 (DP120100895).


\end{document}